\RequirePackage[displaymath]{lineno}
\documentclass[a4paper,11pt]{article}
\pdfoutput=1 

\usepackage{jcappub}
\usepackage[T1]{fontenc} % if needed
\usepackage{aas_macros}
\usepackage{tablefootnote}
\usepackage[displaymath]{lineno}
\usepackage{subfigure}
\usepackage{multirow}
\usepackage{amsmath}

\notoc

\def\gsim{\:\raisebox{-0.5ex}{$\stackrel{\textstyle>}{\sim}$}\:}

\newcommand{\GEV}{\ensuremath{\,\textnormal{GeV}}}
\newcommand{\TEV}{\ensuremath{\,\textnormal{TeV}}}

\begin{document}

\title{Confronting recent AMS-02 positron fraction and  Fermi-LAT Extragalactic $\gamma$-ray  Background measurements with gravitino dark matter}

% Author list

\author[a]{Edson~Carquin,}
\author[a]{Marco~A.~D\'\i az,}
\author[a,b]{Germ\'an~A.~G\'omez-Vargas,}
\author[c]{Boris~Panes,}
\author[a]{Nicol\'as Viaux}

\affiliation[a]{Instituto de Fis\'ica, Pontificia Universidad Cat\'olica de Chile, Avenida Vicu\~na Mackenna 4860, Santiago, Chile}
\affiliation[b]{Istituto Nazionale di Fisica Nucleare, Sezione di Roma ``Tor Vergata'', I-00133 Roma, Italy}
\affiliation[c]{Instituto de F\'\i sica, Universidade de S\~ao Paulo, R. do Mat\~ao 187, S\~ao Paulo, SP, 05508-900, Brazil }

\emailAdd{edson.carquin@fis.puc.cl}
\emailAdd{mad@susy.fis.puc.cl}
\emailAdd{ggomezv@uc.cl}
\emailAdd{bapanes@if.usp.br}
\emailAdd{nviaux@fis.puc.cl}

%\pacs{95.35.+d,95.30.Cq,98.35.Gi}

\abstract{ Recent positron flux fraction measurements in cosmic-rays (CR) made by the AMS-02 detector confirm and extend the evidence on the existence of a new (yet unknown) source of high energy electrons and positrons. We test the gravitino dark matter of bilinear R-parity violating supersymmetric models as this electrons/positrons source.  Being a long lived weak-interacting and spin 3/2 particle, it offers several particularities which makes it an attractive dark matter candidate.  We compute the electron, positron and $\gamma$-ray\  fluxes produced by each gravitino decay channel as it would be detected at the Earth's position. Combining the flux from the different decay modes we are able to reproduce AMS-02 measurements of the positron fraction, as well as the electron and positron fluxes, with a gravitino dark matter mass in the range $1-3$ TeV and lifetime of $\sim 1.0-0.7\times 10^{26}$ s. The high statistics measurement of electron and positron fluxes, and the flattening in the behaviour of the positron fraction recently found by AMS-02 allow us to determine that the preferred gravitino decaying mode by the fit is $W^{\pm}\tau^{\mp}$, unlike previous analyses. Then we study the viability of these scenarios through their implication in $\gamma$-ray observations. For this we use the Extragalactic $\gamma$-ray Background recently reported by the {\it Fermi}-LAT Collaboration and a state-of-the-art model of its known contributors. Based on the $\gamma$-ray analysis we exclude the gravitino parameter space which provides an acceptable explanation of the AMS-02 data. Therefore, we conclude that the gravitino of bilinear R-parity violating models is ruled out as the unique primary source of electrons and positrons needed to explain the rise in the positron fraction.
} 

%\mathbf{on cosmic rays, }


\maketitle
%\flushbottom

\newpage
%%%%%%%%%%%%%%%%%%%%%%%%%%%%%%%%%%%%%%%%%%%%%%%%%%%%%%%%%%%%%%%%%%%%
\section{Introduction}
%%%%%%%%%%%%%%%%%%%%%%%%%%%%%%%%%%%%%%%%%%%%%%%%%%%%%%%%%%%%%%%%%%

%Over the last eight decades vast and compelling evidence of the dark matter (DM) existence at different scales in the universe have been accumulated~\cite{2012JMPh....3.1152R}. However, its fundamental nature have remained as one of the biggest mysteries of modern science \cite{Bertone09,Munoz:2003gx}. The study and characterisation of different cosmic ray (CR) species arriving at the Earth could reveal the DM nature. DM decays or annihilations might inject Standard Model (SM) particles in the interstellar medium (ISM) that will appear as exotic contributions in CR measurements, challenging their interpretation in terms of known astrophysical phenomena. This is the reason why 

 There is vast and compelling evidence of the existence of dark matter (DM), ranging from galactic to cosmological scales~\cite{2012JMPh....3.1152R}. Nevertheless, its fundamental nature remains as a mystery~\cite{Bertone09,Munoz:2003gx}. An indirect way to detect DM at Earth could be by measuring its decay or annihilation products (i.e. Standard Model (SM) particles), which might be injected in the interstellar medium (ISM) by these processes. Thus, the study and characterisation of different CR species may shed light on the DM nature. In this respect, the unexpected increase in the positron fraction ($e^+/(e^+ + e^-)$) at energies above 10 GeV initially detected by PAMELA \cite{Adriani:2008zr}, confirmed by the {\it Fermi}-Large Area Telescope (Fermi-LAT) \cite{FermiPositron}, and measured with precision by the Alpha Magnetic Spectrometer (AMS) \cite{Aguilar:2013qda} attracted a lot of attention from the astro-particles community causing a spate of works trying to explain this anomaly with by-products of DM annihilation or decay, see e.g.~\cite{Feng:2013zca,Bergstrom:2013jra,Ibarra:2013cra,Choi:2013oaa,Hryczuk:2014hpa,Baek:2014awa,Zhao:2014nsa,Cirelli:2008pk,Ibarra:2013zia,Dev:2013hka,Ibe:2013jya,Baek:2014goa}. However, the interpretation of these leptonic CR measurements in terms of DM is not straightforward as the models of DM able to explain the rise in the positron fraction, usually, overpredicts the anti-proton and $\gamma$-ray observations, see e.g. \cite{ArkaniHamed:2008qn,Ibarra:2009dr,Ibe:2013nka,Delahaye:2013yqa,2012PhRvD..86h3506C,Ando:2015qda}\footnote{Note that decaying DM particles could affect the physics of dense stars implying strong constraints on the DM properties~\cite{2014arXiv1403.6111P}. These constraints are avoided by the gravitino because its elastic scattering cross section on nuclei is proportional to $1/M_{Pl}^4$~\cite{Grefe:2011dp}, therefore there is no chance that they can be trapped in the core of dense stars. In particular, antideuterons produced by gravitino decays were recently studied in~\cite{Grefe:2015jva,Dal:2014nda,Monteux:2014tia}.}. Astrophysical hypotheses explaining the positron rise, as for instance nearby pulsars or spalls from primary CRs\footnote{ The most common primary CR particle is the ubiquitous proton or hydrogen nucleus. 95\% of all CRs are protons, 4\% are helium nuclei, and the 1\% balance is made up of nuclei from other stellar-synthesized elements up to iron. Their principal Galactic source are: supernova remnants, neutrons stars, black holes and possibly DM} interacting with the ISM, work well reproducing the data, see e.g.~\cite{DiMauro:2014iia,Hooper:2008kg}. Nevertheless, these proposals have their own difficulties in each case \cite{Igor}.

The AMS Collaboration has extended the measurement of both, positron fraction and positron flux to 500 GeV, as well as the electron spectrum to 700 GeV~\cite{AMS,AMS1}. The main highlights of these measurements are: i) the positron fraction  rise starts to flatten at energies above $\sim 200$ GeV \cite{AMS}, and ii) precision measurements of electron and positron fluxes allow determination of different behaviour of their spectral indexes as the energy increases \cite{AMS1}. Therefore, in light of the latest AMS-02 release \cite{AMS,AMS1} we can better determine the properties of the cosmic source responsible for the rise in the positron fraction.

A minimal model to characterise the electron and positron fluxes, and the anomalous behaviour of the positron fraction can be written as \cite{Aguilar:2013qda,AMS,AMS1},

\begin{eqnarray}
 \Phi_{e^-}(E)&=&C_{e^-}(E/1~\mbox{GeV})^{-\gamma_{e^-}}+C_s (E/1~\mbox{GeV})^{-\gamma_s}e^{-E/E_s}, \nonumber \\
 \Phi_{e^+}(E)&=&C_{e^+}(E/1~\mbox{GeV})^{-\gamma_{e^+}}+C_s (E/1~\mbox{GeV})^{-\gamma_s}e^{-E/E_s},
\label{eq:AMS_ansatz}
\end{eqnarray}

\noindent where each equation is composed of an astrophysical background term (modelled with a power-law spectrum), and a single source term (modelled with a power-law spectrum with an exponential cutoff energy) which is common to the electron and positron flux parameterisations. The  astrophysical background term in the electron flux attempt to model primary and secondary\footnote{It is due to spallation in the ISM owing to the impacts of primary cosmic rays off ISM nuclei.} production, while in the positron flux equation represents only secondary production. This minimal model describe in good agreement the positron fraction and the combined electron and positron flux~\cite{Aguilar:2013qda}. Also, this parametrisation reveals and characterises the spectrum and intensity of a common source of primary electrons and positrons required to fit the high energy end of the spectra. 

 %Therefore, the data can be described by a common source of primary electrons and positrons~\cite{Aguilar:2013qda}.

%\noindent where the numerical values of the parameters $C_{e^-}$, $C_{e^+}$, $\gamma_{e^-}$, $\gamma_{e^+}$, $C_s$, $\gamma_s$ and $E_s$ are not initially fixed. Notice that each flux is determined by the sum of a power-law spectrum, which dominates at low energies and a symmetric contribution (the source) which is used to tune the high energy region of the positron fraction and the electron-positron flux. The behaviour of the data at low energies requires that $C_{e^-}\gg C_{e^+}$. With this minimal model satisfactory fits imply that the positron rise can be described by a common source of primary electrons and positrons.

%It turns out that the positron fraction depends on just five independent combinations of these parameters, whose updated values can be found fitting the current data. 

In this work we assume the single common source term in eq.~\ref{eq:AMS_ansatz} consisting of decaying DM particles, which preserves CP symmetry, and thus contribute an equal amount to the electron and positron flux. In particular, we study a decaying gravitino DM scenario, which is possible in the context of R-parity violating (RpV) supersymmetric (SUSY) models, see e.g.~\cite{Takayama:2000uz,Buchmuller:2007ui,Grefe:2008zz,Choi:2010jt,Restrepo:2011rj,Ibe:2013nka,Diaz:2011pc,Cottin:2014cca,Bomark:2009zm,Bajc:2010qj} \footnote{Unlike~\cite{Buchmuller:2007ui,Choi:2010jt,Restrepo:2011rj,Diaz:2011pc,Cottin:2014cca} we focus on gravitino DM with mass above 1 TeV motivated by the flattening of the positron fraction spectrum at $\sim$ 200 GeV. The study of the consistency between this gravitino mass scale and other observables, as neutrino physics, is beyond the scope of this work.}. We assume that the relic density produced by our gravitino DM, is consistent with the observations, as already has been shown in several places
\cite{Fayet:1981sq,Giudice:1999am,Bolz:2000fu,Pradler:2006qh,Rychkov:2007uq}. It is worth noting that these works assume that the gravitino is sufficiently long-lived in order to maintain its comoving density from the period of decoupling until the present. In fact, since RpV terms are in general quite small and the gravitino interactions are suppressed by the Planck mass, the latter condition is naturally obtained \cite{Buchmuller:2007ui}.

In SUSY with RpV, the gravitino decay channels are fixed (see for instance \cite{Takayama:2000uz,Moreau:2001sr,Buchmuller:2007ui,Ishiwata:2008cu,Grefe:2008zz}). In this work we just consider the primary  {\it two body} decay channels which could produce electrons and positrons in the final state. In models with trilinear RpV two body decay channels are possible at one-loop level, but for heavy gravitinos the behaviour of the total decay rate is dominated by three-body decay channels at tree level~\cite{Bomark:2009zm,Bajc:2010qj}. Therefore, we implicitly assume negligible R-parity violation due to trilinear couplings, so our theoretical framework are the bilinear RpV models. For several values of the gravitino mass, final gravitino decay products are computed using Pythia 8.185 \cite{PYTHIA8} and propagated from regions with non-zero DM densities towards the Earth. The obtained spectrum produced by each of the different decay channels are combined in order to fit the most recent AMS-02 data. From the results of this fitting procedure, we can extract the best values for the corresponding branching ratios (BR). We then compute the corresponding  diffuse $\gamma$-ray emission produced by the obtained gravitino parameters at high Galactic latitudes, which turns out to significantly exceeds the flux of the extragalactic $\gamma$-ray background (EGB) recently measured using the {\it Fermi}-LAT~\cite{Ackermann:2014usa}\footnote{ The EGB comprises detected point sources, all extragalactic emissions too faint or too diffuse to be resolved, and any residual Galactic foregrounds that are approximately isotropic. Therefore, the EGB could contains emission from gravitino DM decays of extragalactic and Galactic origin, it has been used to set stringent constrains on DM lifetime, see \cite{2012PhRvD..86h3506C} and references therein.}. Finally, we compute the residual upper limit flux of the EGB removing its known contributors, in particular the integrated emission from blazars as modelled in~\cite{AjelloBlazars}, and then compare it to the prompt gravitino-induced  $\gamma$-ray flux to obtain conservative limits on the gravitino lifetime that can be directly compared to the lifetime required to account for the AMS-02 leptonic measurements. 

This work extends and update previous studies in several respects. As we try to be as model independent as possible, we are not fixing the branching ratios for the different gravitino decay channels, neither the flavour structure in our theory, but rather, this is the outcome of the fits to the AMS-02 data. Indeed, we present for the first time a fit of the branching ratios for the different gravitino decay channels to data, finding that the best fit parameters are compatible with the theoretical expectation for gravitinos with masses in the range between 750 and 3000 GeV, $W^{\pm}l^{\mp}_i$:$Z\nu_i$:$H\nu_i$ in proportion 2:1:1~\cite{Ishiwata:2008cu,Grefe:2011dp,Delahaye:2013yqa}. Furthermore, we are using the most up to date results from AMS-02 and {\it Fermi}-LAT collaborations, which considerably extends previous measurements. It allows a more precise characterisation of the source responsible for the positron fraction excess and reduce the window for exotic contributors to the $\gamma$-ray sky. In particular, the flattening in the behaviour of the positron fraction at about 200 GeV and the high statistics measurement of electron and positron fluxes allow us to determine the mass scale of the gravitino DM ($\sim 1$ TeV ) and  that the gravitino decaying mode preferred by AMS-02 data is $W^{\pm}\tau^{\mp}$. The last differs from previous analyses where the favoured channel is $W^{\pm}\mu^{\mp}$~\cite{Ibarra:2009dr,Ibe:2013nka},  we discuss further on this difference in section \ref{sec:fit}. It is worth noting that we only use in the fit the positron fraction and positron flux measurements  in order to increases the reduced $\chi^2$ as the number of degrees of freedom is reduced. We do this to emulate the effect of assuming a covariance matrix correlating the three datasets, which is beyond the scope of this work. We ruled out the hypothesis of gravitino DM in bilinear RpV models as the sole explanation for the rise in the positron fraction using the latest determination of the EGB~\cite{Ackermann:2014usa}.

The paper is organised as follows: In section 2 we introduce the primary gravitino decay channels which are available in SUSY with bilinear RpV models. In section 3 we consider the propagation of electrons and positrons produced by the gravitino decays through the ISM in three different propagation model scenarios. Afterwards we fit the most recent leptonic data published by the AMS Collaboration using a model that consider a power-law background plus a source given by the gravitino decays\footnote{As already mentioned, we only use positron flux and positron fraction measurements in the fit, then we use the obtained gravitino and background parameters to check visually a reasonable agreement of the electron flux with the measurement.}. In section 4 we use the results obtained from the fit to AMS-02 data in order to determine the $\gamma$-ray signal from Galactic gravitino DM decay at high latitudes and we confront these results with the new {\it Fermi}-LAT EGB measurement and latest modelling of integrated blazar emission. In section 5  we use the EGB measurement to constrain the gravitino DM models that provide an acceptable explanation to the positron rise. Finally, in section 6 we conclude.

%%%%%%%%%%%%%%%%%%%%%%%%%%%%%%%%%%%%%%%%%%%%%%%%%%%%%%%%%%%%%%%%%%%%
\section{Gravitino decay modes}
\label{gdecay}
%%%%%%%%%%%%%%%%%%%%%%%%%%%%%%%%%%%%%%%%%%%%%%%%%%%%%%%%%%%%%%%%%%%%

%Our decaying DM candidate consists of a heavy gravitino with bilinear R-Parity violating couplings to SM particles and the neutralinos/charginos.
Our decaying DM candidate consists of a heavy gravitino that decay through RpV terms.  As already pointed out, we will study gravitino LSP. We consider $m_{3/2}\gsim 1$ TeV motivated by the most recent positron fraction measurement presented by the AMS Collaboration. In this regime the main dacay modes of the gravitino are the following~\cite{Ishiwata:2008cu,Grefe:2008zz},
%In this model, although the gravitino is heavy ($\gsim$ 1 TeV), it is the LSP, and then it can only decay to SM particles. The rest of the SUSY particles in the spectrum are even heavier than the gravitino and are in general safe from current LHC constraints. We consider in this work the following decay modes of the gravitino \cite{Ishiwata:2008cu,Grefe:2008zz},

\begin{equation}
  \Psi_{3/2} \rightarrow W^{\pm}l^{\mp}_i,\ Z\nu_i,\ H\nu_i
\label{eq:decaychannels}
\end{equation}

\noindent where $i$ is the lepton family index\footnote{The Gravitino decay to $\gamma\nu_i$ is strongly suppressed in the limit where $m_{3/2}$ is heavier than the weak gauge bosons and will not be further considered in this work, also we do not consider decays of the gravitino involving the extra Higgs bosons predicted in SUSY theories, because they are generally heavier than the LSP~\cite{Ishiwata:2009vx,Grefe:2011dp,Delahaye:2013yqa}. }. It has been shown in \cite{Delahaye:2013yqa} that in the limit of gravitino masses above 1 TeV, the decay widths of the channels listed in \ref{eq:decaychannels}, reach their asymptotic limit,
\begin{eqnarray}
\Gamma(\psi_{3/2}\rightarrow W^{\pm}l^{\mp}_i) & \simeq & \frac{m_{3/2}^{3}}{192\pi M_{Pl}^{2}}\left(\frac{\Lambda_i}{\mu v}\right)^{2},\\
\Gamma(\psi_{3/2}\rightarrow Z\nu_i) & \simeq & \frac{m_{3/2}^{3}}{384\pi M_{Pl}^{2}}\left(\frac{\Lambda_i}{\mu v}\right)^{2}, \\
\Gamma(\psi_{3/2}\rightarrow H\nu_i) & \simeq & \frac{m_{3/2}^{3}}{384\pi M_{Pl}^{2}} \left(\frac{\Lambda_i}{\mu v}\right)^{2}.
\end{eqnarray}

\noindent Where $\Lambda_i$ are the effective bilinear RpV terms~\cite{Hirsch:2000ef}, $\mu$ is the higgsino mass and $v$ the vacuum expectation value. The corresponding branching ratios become independent of the size of R-parity violation, while their relative sizes converge to the proportion 2:1:1, for  $W^{\pm}l^{\mp}_i$, $Z\nu_i$, and $H\nu_i$, respectively. From now on we generically refer to these final states as $\lambda l_i$. Since we are indeed interested in the heavy gravitino regime, where the hierarchy of the branching ratios for the different decay channels holds, we assume this as our benchmark scenario.%, against which we will confront the result of our analysis of the new AMS data.
%$Wl=\sum_i W^{\pm}l_i^{\mp}$ and $Z(h)\nu=\sum_{\substack{i}} Z(h)\nu_i$

\begin{figure}[htb]
  \subfigure[]{
   \includegraphics[width=0.5\linewidth]{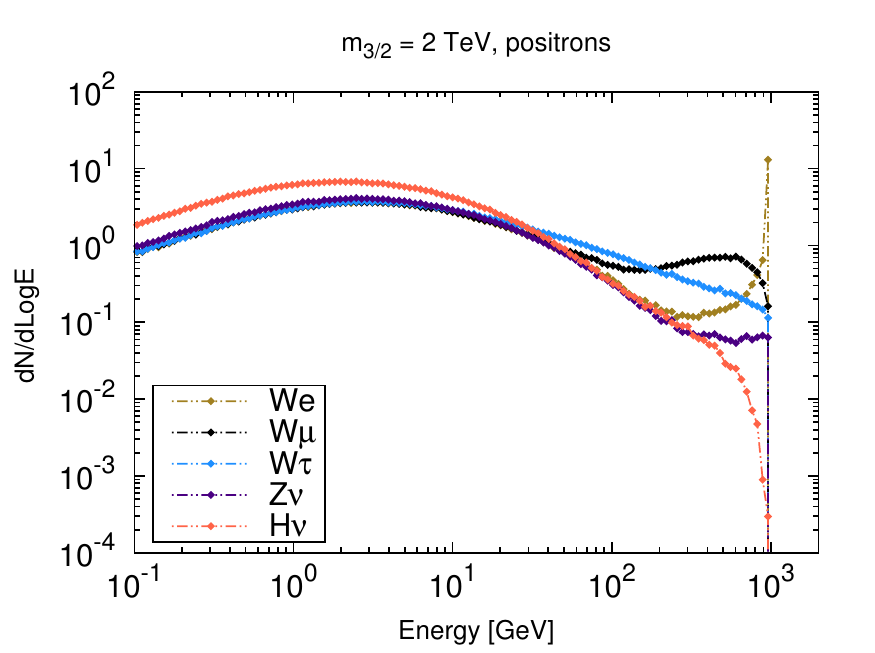}
  }
  \subfigure[]{
    \includegraphics[width=0.5\linewidth]{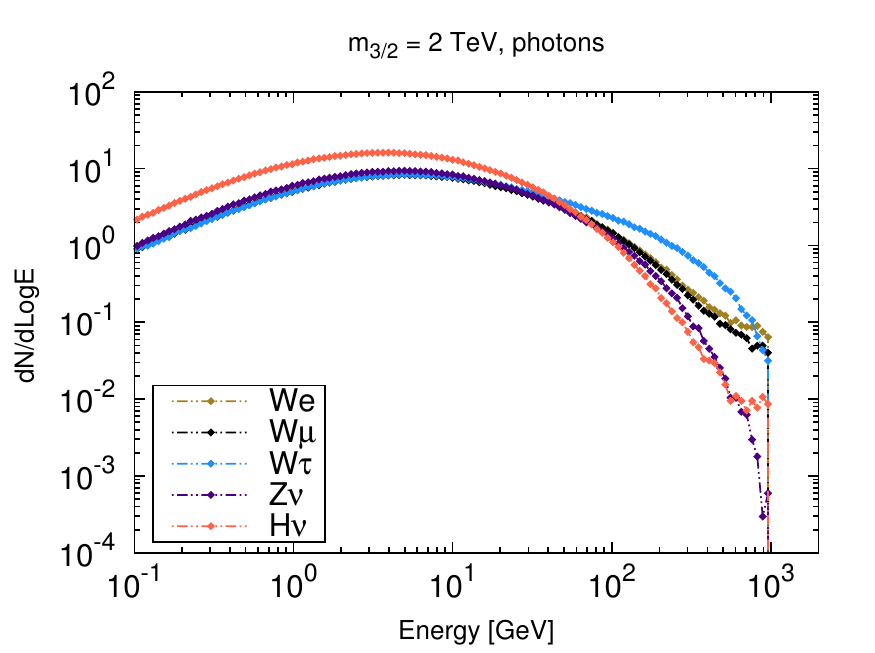}
  }
  \caption{The positron (a) and photon (b) energy spectrum produced by the different gravitino decay modes as a function of energy. The electron distribution is equivalent to the positron case.}
\label{fig:inspectra}
\end{figure}

 We use Pythia 8.185 \cite{PYTHIA8} to compute the number of final state electrons, positrons and photons, as well as their 4-momenta, produced by each single gravitino decay mode. In fig.~\ref{fig:inspectra} we show the resulting energy spectra. 
%The final states are fully simulated including the decay kinematics and the corresponding hadronic showers to take into account coloured final states, including final state radiation

%%%%%%%%%%%%%%%%%%%%%%%%%%%%%%%%%%%%%%%%%%%%%%%%%%%%%%%%%%%%%%%%%%%%
\section{AMS-02 positron excess and gravitino dark matter}
\label{sec:AMS}
%%%%%%%%%%%%%%%%%%%%%%%%%%%%%%%%%%%%%%%%%%%%%%%%%%%%%%%%%%%%%%%%%%%%
\subsection{Cosmic ray propagation}
\label{sec:propagation}
%%%%%%%%%%%%%%%%%%%%%%%%%%%%%%%%%%%%%%%%%%%%%%%%%%%%%%%%%%%%%%%%%%%%

Once a gravitino decay in the halo of the Milky Way (MW), some of the resulting charged particles will propagate through the ISM towards the Earth, where they might be detected. Theoretically, the propagation of electrons and positrons in the MW can be treated as a diffusive process \cite{Strong:2007nh}. If we assume that this is a {\it steady state process}, we can estimate the electron and positron density per unit energy at position $r$ (in Galactic coordinates) i.e. $f(E,r)$, by solving the following diffusion equation,

\begin{equation}
-\vec{\nabla}\cdot\left( \mathcal{K}(E,r) \vec{\nabla} f \right)- \frac{\partial}{\partial E}\left( b(E,r) f \right) = Q(E,r),
\label{diff}
\end{equation}

\noindent where $\mathcal{K}(E,r)=\mathcal{K}_{0}(E/GeV)^{\delta}$ is the diffusion coefficient that account for the transport of electrons and positrons through the Galactic magnetic fields, $b(E,r)$ is the energy loss coefficient due to the different types of interactions of electrons and positrons with the ISM, like synchrotron radiation, bremsstrahlung and inverse Compton scattering (ICS), and $Q$ is the {\it source} term which model how electrons and positrons are injected in the ISM, in our case, this is due to gravitino decays, and is given for each $\lambda l_{i}$ channel as:

\begin{equation}
Q(E,r)=\left(  \frac{\rho (r)}{m_{3/2}} \right) \frac{1}{\tau_{3/2}}  \frac{dN^{3/2}_{\lambda l_i\rightarrow e^\pm}}{dE}.
\label{sourcet}
\end{equation}

Replacing this expression for $Q(E,r)$ in eq. 3.1, we can determine the flux of electrons and positrons as a function of the energy $E$ and position $r$, i.e. $d\Phi_{\lambda l_i\rightarrow e^{\pm}}^{3/2}/dE\, (E,r) = v_{e^\pm} f(E,r)/4\pi$, using the following equation \cite{cirelli}\footnote{In \cite{Delh2008}  the eq.~(\ref{diff}) is resolved for the MW configuration.}: 

\begin{equation}
\frac{d\Phi_{\lambda l_i\rightarrow e^{\pm}}^{3/2}}{dE}(E,r) =
\frac{v_{e^\pm}}{4\pi \, b(E,r)}\phantom{\frac12 }\left(\frac{\rho(r)}{m_{3/2}}\right) \frac{1}{\tau_{3/2}} \int_E^{M_{\rm DM}/2} dE_{\rm s} \frac{dN^{3/2}_{\lambda l_i\rightarrow e^\pm}}{dE}(E_{\rm s}) \,{I}(E,E_{\rm s},r),
\end{equation}

\noindent where $v_{e^\pm}$ is the velocity of the electrons and positrons, $E_{\rm s}$ is their energy at the source and $I(E,E_{\rm s},r)$ are the {\em generalized halo functions} (GHF), which are essentially the Green functions connecting a source with a fixed energy $E_{\rm s}$ to a given energy $E$ (see \cite{Delh2008} for more details of the GHF). In this work we use the Navarro, Frenk and White (NFW) \cite{nfw} density profile

\begin{linenomath}
\begin{equation}
 \rho_{\text{NFW}}(r)=\rho_{s}\frac{r_s}{r}\left( 1 + \frac{r}{r_s} \right)^{-2},
\end{equation}
\end{linenomath}

\noindent where we adopt $r_s=24.42\,$kpc and $\rho_s=0.184\,\text{GeV}\,\text{cm}^{-3}$ following \cite{cirelli}, where at Earth position $r_{\odot}=8.33\, {\rm kpc}$ \cite{Gillessen09} the local DM density is assumed to be  $\rho _\odot =0.3$ GeV cm$^{-3}$. The diffusion model is defined by $\mathcal{K}_{0}$, ${\delta}$ and the half-thickness $L$ of the diffusion zone\footnote{It is define as a solid flat cylinder with height $2L$ in the $z$ direction and radius $R=20$kpc in the radial direction.}. We use the set of propagation models called: MIN, MED, MAX described in \cite{Delh2008}. The MIN (MAX) model minimizes (maximizes) the $\overline{p}$ flux at Earth and are in agreement with the measurements of the boron-to-carbon (B/C) ratio at the Earth's position \cite{Delh2008}. This models contains parameters for the propagation model that achieve an acceptable configuration for CRs propagation in the MW. The use of these models are aimed at estimating the impact of the uncertainties inherent to electrons and positrons propagating in the ISM. Note that MIN, MED, and MAX scenarios do not test the complete range of systematics related to the modelling of CRs propagation  \cite{Delh2008}.  Furthermore, the  MIN scenario is strongly disfavoured by CR and gamma-ray data \cite{Lavalle:2014kca,Giesen:2015ufa}.

\begin{figure}[t]
 \subfigure[]{
  \includegraphics[width=0.5\linewidth]{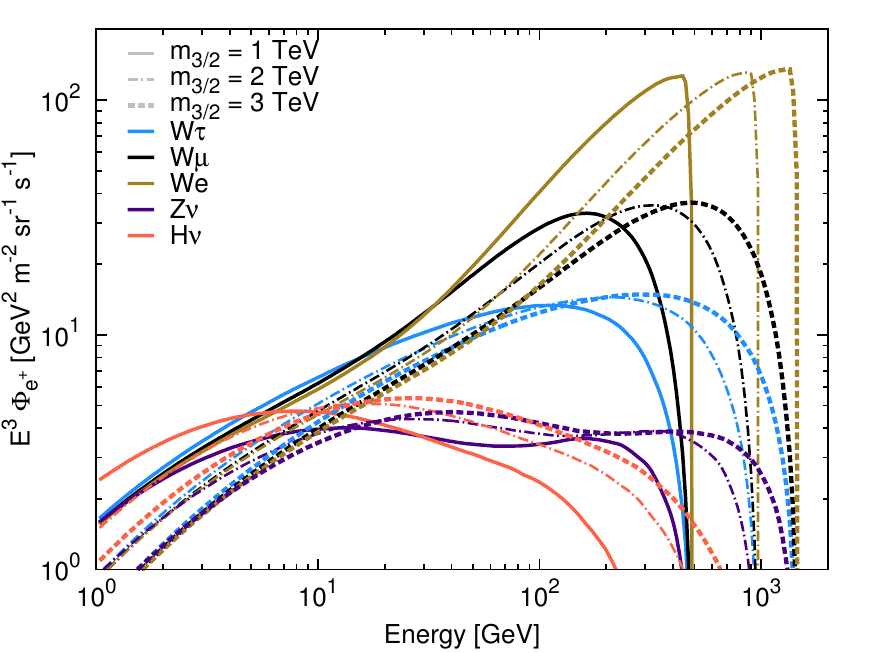}
  }
  \subfigure[]{
  \includegraphics[width=0.5\linewidth]{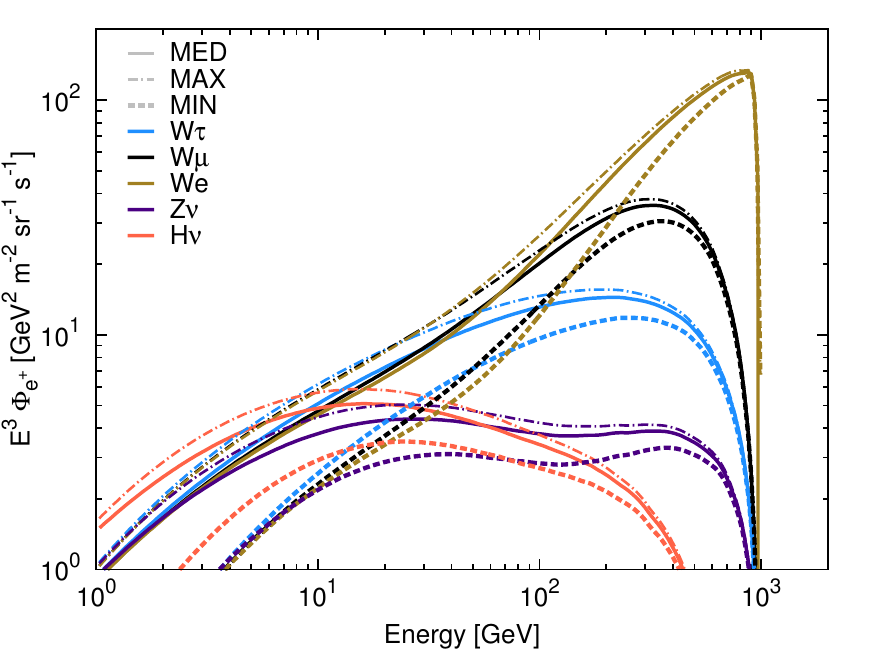}
  }
 %  \centering
  %  \includegraphics[width=0.8\linewidth]{Flux_DM_M_123TeV_Nicolas.pdf}
 %\includegraphics[width=0.8\linewidth]{Flux_DM_M_123TeV_Nicolas.eps}
  \caption{a) Propagated flux at Earth using the MED scenario for a gravitino mass of 1 TeV, 2 TeV and 3 TeV with fixed $\tau _{3/2}=10^{26}s$, different colours indicates the different channels of the gravitino decay, and we are assuming only one decay mode at a time, with a $BR_{\lambda l_{i}}$ equal to one. b) Propagated flux at Earth using the MIN, MED, MAX scenarios for a gravitino mass of 2 TeV. We note that MIN scenario is strongly disfavoured by recent CR and gamma-ray data \cite{Lavalle:2014kca,Giesen:2015ufa}. It is worth noting that MIN and MAX scenarios do not bracket all the uncertainty due to propagation of CRs in the ISM.}
  \label{fig:data}
\end{figure}

In panel a) of fig.~\ref{fig:data} we show the propagated electron spectra at the Earth's position for different gravitino masses using MED propagation model. We show the propagated spectra of electrons from gravitino decays with masses of $1$, $2$ and $3$ TeV. We use this set of curves in the following section as the inputs to the fit of the AMS-02 data.  In panel b) of figure~\ref{fig:data} we present the effect on the gravitino-induced electron flux at Earth's position due to the three different propagation models MIN, MED, and MAX. We see in this figure that the effect of different propagation scenarios is less than one order of magnitude, but in any case as we will see later the results of the fit are barely affected by the choose of the propagation model.

%%%%%%%%%%%%%%%%%%%%%%%%%%%%%%%%%%%%%%%%%%%%%%%%%%%%%%%%%%%%%%%%%%%%
\subsection{Determination of gravitino parameters from AMS-02 data }
\label{sec:fit}
%%%%%%%%%%%%%%%%%%%%%%%%%%%%%%%%%%%%%%%%%%%%%%%%%%%%%%%%%%%%%%%%%%%%
In this section we consider a decaying gravitino scenario as describing the anomalous behaviours of the electron and positron fluxes, and the positron fraction, all of them recently measured with high precision by the AMS-02 experiment \cite{AMS,AMS1}. Since the three measurements are not fully independent among them, correlations should be included in the fit, for instance by using a complete covariance matrix. Unfortunately, the computation of this matrix is beyond the scope of this work. Instead of a full statistical analysis assuming a covariance matrix we make an approximation in the fit computation. In practice, we just take into account two of the three datasets plus a diagonal covariance matrix, which can be computed using the standard deviations released by the AMS Collaboration. In this way we reduce the artificial over constraining of the free parameters that we would obtain from the fit using the three data sets disregarding correlations among them. 

From the three possible combinations of datasets we consider the positron flux and positron fraction, excluding the electron flux. By using these datasets it is possible to determine unambiguously both, the best fit values of the free parameters and their error intervals. Similar results, both qualitatively and quantitatively, are obtained at using the electron flux and positron fraction datasets. For the remaining case, electron and positron fluxes, it turns out that the normalisation and spectral index of positron flux, $C_{e^+}$ and $\gamma_{e^+}$ , cannot be determined. It is so because the high energy part of the measured positron flux can be well fitted alone by the gravitino-induced positron emission, therefore we just need a mild contribution of the positron flux at low energies that can be arranged by using unconstrained combinations of $C_{e^+}$ and $\gamma_{e+}$. 

%In this section, we consider a decaying gravitino scenario as describing the anomalous behaviours of the electron and positron flux, and the positron fraction, all of them recently measured with high precision by the AMS-02 experiment \cite{AMS,AMS1}. In practice we {\bf only fit the positron flux above 10 GeV and the positron fraction above 1 GeV,  as a proxy mimic the correlations among the three measurements. Using two measurements we increase the reduced $\chi^2$ with respect to using the three correlated measurements as the number of degrees of freedom is reduced.  The reason to exclude the electron flux from the fit is that this is less sensitive to changes in the single-source term (see eq.~\ref{eq:AMS_ansatz}) than the positron flux. In other words, following eq.~\ref{eq:AMS_ansatz}, as the ratio $C_{s}/C_{e^+}$ is one order of magnitud larger than $C_{s}/C_{e^-}$ \cite{Aguilar:2013qda,AMS} the behaviour of the single-source term can be better pin down adding the positron flux information to the positron fraction, than the electron flux measurement.} We do not consider the positron flux below 10 GeV because the features in which we are interested on appear at energies larger than 20 GeV. Besides this, AMS-02 data below 10 GeV is strongly affected by the Solar activity~\cite{Mosk98}. 

Motivated by eq.~\ref{eq:AMS_ansatz} we model the electron and positron flux by considering the sum of a non-symmetric power law background plus the contribution of a symmetric source, in our case, produced by gravitino decays in the the Galactic halo, which is assumed to make the whole DM density in the MW. The latter assumption can be relaxed when we consider that a different value of the average density can be absorbed by a redefinition of the gravitino lifetime. The only constraint would come from the metastability condition, that roughly speaking requires $\tau_{3/2}> 10^{17} \text{s}$. The electron and positron fluxes are given by,

\begin{eqnarray}
\Phi_{e^{-}}(E) & = & C_{e\text{\textsuperscript{-}}}E{}^{-\gamma_{e\text{\textsuperscript{-}}}}+\Phi_{e^{-}}^{3/2}(E,m_{3/2},\tau_{3/2},BR_{\lambda l_i}), \nonumber \\
\Phi_{e^{+}}(E) & = & C_{e\text{\textsuperscript{+}}}E{}^{-\gamma_{e\text{\textsuperscript{+}}}}+\Phi_{e^{+}}^{3/2}(E,m_{3/2},\tau_{3/2},BR_{\lambda l_i}). 
\label{eq:fluxes_gravitino}
\end{eqnarray}

The asymmetric contribution to the $e^{\mp}$ flux is determined by the corresponding normalisation factors $C_{e^{\mp}}$ and spectral indexes $\gamma_{e^{\mp}}$. The term $\Phi_{e^{\mp}}^{3/2}(E,m_{3/2},\tau_{3/2},BR_{\lambda l_i})$ corresponds to the symmetric contribution to the $e^{\mp}$ flux arising from the decay of gravitinos. As already mentioned in section \ref{sec:propagation}, the gravitino contribution depends on its mass $m_{3/2}$, lifetime $\tau_{3/2}$ and decay branching ratios $BR_{\lambda l_i}$. Notice that one of the gravitino BRs is not independent because we have assumed that the sum of them must be equal to unity. Apart from the latter restriction, during the fit we consider these variables as free parameters. 

Since the computation of the gravitino contribution to the $e^{\mp}$ flux is quite sensitive to the value of its mass, for simplicity we have considered just three specific cases given by $m_{3/2}^{*}=1,\,2$ and $3\,\TEV$. We will see later that these cases are enough to extract general conclusions from the fit to the data. For each of these masses we compute a set of spectra corresponding to the different decay channels with fixed lifetime $\tau_{3/2}^{*}=10^{26}\, \text{s}$ after propagation through the ISM. These spectra are showed in fig.~\ref{fig:data}. The flux obtained for this specific lifetime can be linearly scaled in order to obtain the flux for any other value of $\tau_{3/2}$. Thus, we model the gravitino contribution in the fit computation with the following expression,

\begin{equation}
\Phi_{e^{\mp}}^{3/2}(E,m_{3/2}^{*},\tau_{3/2},BR_{\lambda l_i})=\frac{\tau_{3/2}^{*}}{\tau_{3/2}}\sum_{\lambda l_i}BR_{\lambda l_i}\Phi_{\lambda l_i\rightarrow e^{\mp}}^{3/2}(E,m_{3/2}^{*},\tau_{3/2}^{*}).
\end{equation}

Finally, in order to determine the best fit values of the free parameters, for each of the $m_{3/2}^{*}$ considered we minimise a $\chi$-squared function, $\chi^{2}(C_{e^{+}},\gamma_{e^{+}},C_{e^{-}},\gamma_{e^{-}},\tau_{3/2},BR_{\lambda l_i})$, by considering the eq. \ref{eq:fluxes_gravitino} and the data points and systematic errors from  positron flux and  positron fraction \cite{Battiston20146,AMS}. Indeed, we sum two $\chi$-squared functions constructed from each data set. We use $48$ points from positron flux ($E\geq10\, GeV$) plus $65$ points from positron fraction ($E\geq1\, GeV)$.

In table \ref{tab:Results-of-AMS} we report the values of the best fit parameters using the reference MED propagation model. Also, we compute the error intervals at $1\sigma$ confidence level for the free parameters following \cite{LLyons1986}. It is worth noting that the lifetimes obtained in the fits are close to the nominal value $\tau_{3/2}^{*}=10^{26}\, \text{s}$ in the three scenarios, while the dominant BRs are given by the channels $\Psi_{3/2}\rightarrow W^\pm\tau^\mp$, $\Psi_{3/2}\rightarrow Z\nu$ and $\Psi_{3/2}\rightarrow W^\pm e^\mp(\mu)^\mp$ in descending order. 

Including the theoretical prediction for the BR proportions 2:1:1 (see section \ref{gdecay}) in the fit we obtain reasonable p-values (p>0.05), but the reduced $\chi$-squared is quite large in comparison to the minimum obtained with free BRs\footnote{The $\Delta \chi^2$ with respect to the minimum turns out to be larger than 11.78, which determines the $1\sigma$ region around the global minimum.}.  The resulting gravitino lifetimes when the theoretical prediction on the BRs is included are lower than in the case with free BRs, they are in the range $5.9 - 6.5 \times 10^{25}$ s, these lifetimes are excluded by the limits from antiproton measurements derived in \cite{Delahaye:2013yqa}. The antiproton limits in \cite{Delahaye:2013yqa} cannot be directly applied to our results with free BRs, which provide better description of the AMS-02 leptonic data, therefore in next sections we use gamma-ray measurements to probe those gravitino scenarios.

 %the best fits of the data considering this restriction have p-values\footnote{The $\Delta \chi^2$ with respect to the minimum turns out to be larger than 11.78, which determines the $1\sigma$ region around the global minimum.} equal to 0.24, 0.55, and 0.85 for 1, 2, and 3 TeV, respectively, with the $W^\pm\tau^\mp$ channel remaining the dominant one.} 
 
\begin{figure}[t!h!]
   \centering
    \includegraphics[width=0.8\linewidth]{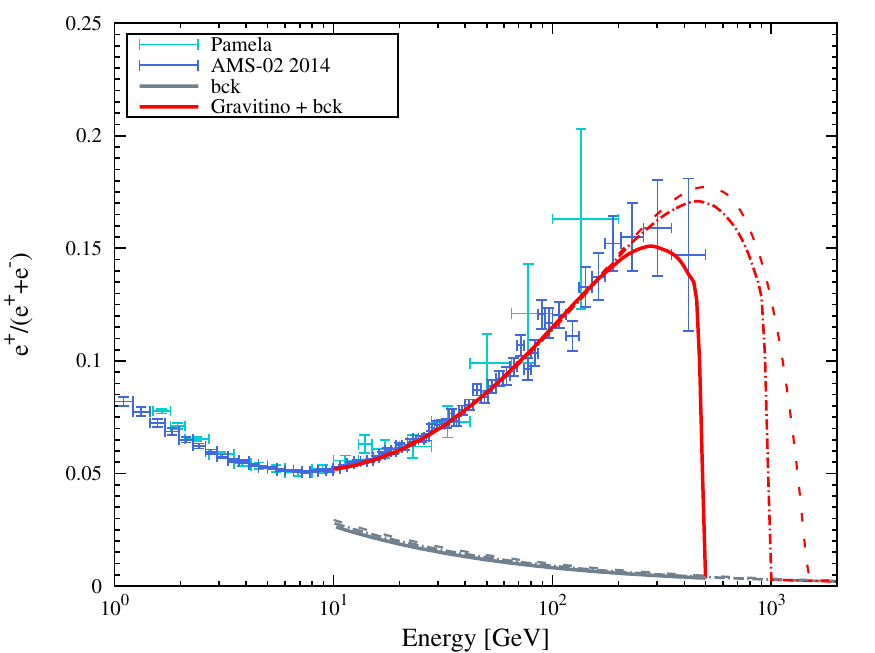}
    \caption{The positron fraction measured by AMS-02 and PAMELA (blue and cyan points respectively), and fits using the gravitino decay products + background for three different values of the gravitino mass (in red), 1 (continuum line), 2 (dot-dashed) and 3 TeV (dashed). The background and gravitino contributions are also shown separately in grey and pale brown respectively.}  
  \label{fig:data1}
\end{figure}

For the fit with free BRs we find that the main gravitino decaying channel is  $W^\pm\tau^\mp$. This result differs from conclusions in \cite{Ibarra:2009dr} and \cite{Ibe:2013nka} where the gravitino decay mainly into the second lepton generation.  In the following we go further on the differences between those analysis and this work that make the $W^\pm \tau^\mp$ channel the preferred by the fit to explain the AMS-02 leptonic data. In \cite{Ibarra:2009dr} the PAMELA positron fraction \cite{Adriani:2008zr} and {\it Fermi}-LAT electron plus positron flux \cite{2009PhRvL.102r1101A} are interpreted in terms of decaying DM. In particular one important difference between the data used in \cite{Ibarra:2009dr} and this work is that the positron fraction measured by PAMELA is harder with respect to the fraction measured by AMS-02, see figure \ref{fig:data1}. It favours lighter families, as can be seen in figure \ref{fig:data}; i.e. for lighter families, both, electron and positron fluxes become harder. Also, it is important to note that  \cite{Ibarra:2009dr} only studies pure DM decaying channels unlike our approach where we combine all of them. The other work \cite{Ibe:2013nka} concludes that the $W^\pm \mu^\mp$ is the favoured gravitino decaying channel by comparing the positron fraction yield by a benchmark point using two different pure decay channels,  $W^\pm e^\mp$ and  $W^\pm \mu^\mp$, to the AMS-02 data.  In \cite{Ibe:2013nka} there is no fit to AMS-02 data, and the electron and positron background fluxes are determined using the electron flux measured by PAMELA \cite{2011PhRvL.106t1101A}. Therefore, the inclusion of high statistic new data and the reduction of error bars provide us new insight in the determination of the source responsible for the positron fraction rise, making plausible our new result.

%It is worth noting that the BRs structure found in our fit is consistent, within $1\sigma$ level, with the theoretical expectation for gravitinos with mass in the TeV range, $W^{\pm}l^{\mp}_i$:$Z\nu_i$:$H\nu_i$ in proportion 2:1:1~\cite{Ishiwata:2008cu,Grefe:2011dp,Delahaye:2013yqa}.  %If we impose this relation in the BR distribution, the minimum $\chi^2$ found correspond to ($W\mu$ + $W\tau$):$Z\nu$:$H\nu$ in proportion 2:1:1.
% Although we perform the fit using MIN, MED, and MAX propagation parameters, in table~\ref{tab:Results-of-AMS} we present the results for MED, since these results are compatible with MIN and MAX scenarios. 

\begin{figure}[h!]
  \subfigure[]{
  \includegraphics[width=0.5\linewidth]{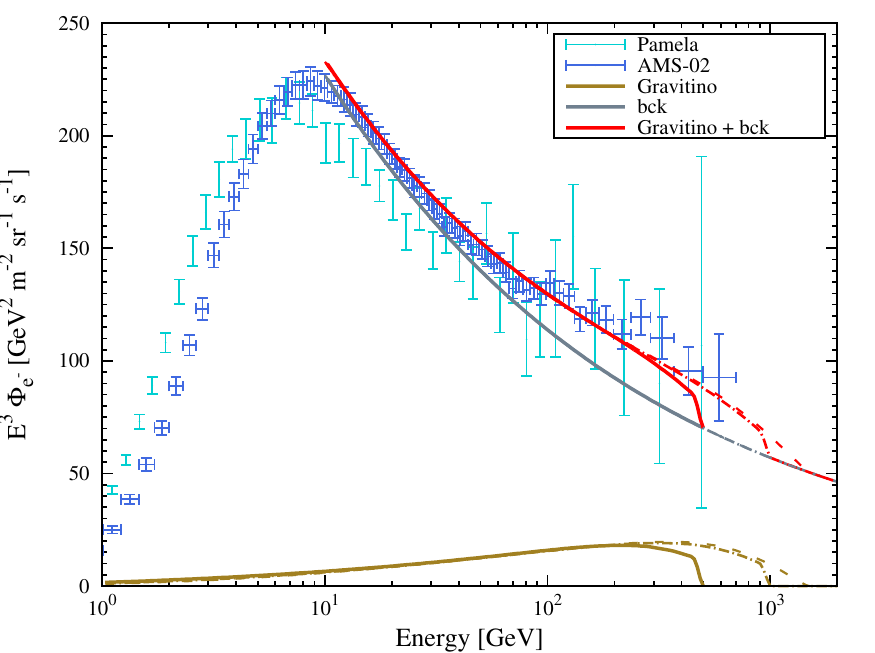}
  }
  \subfigure[]{
  \includegraphics[width=0.5\linewidth]{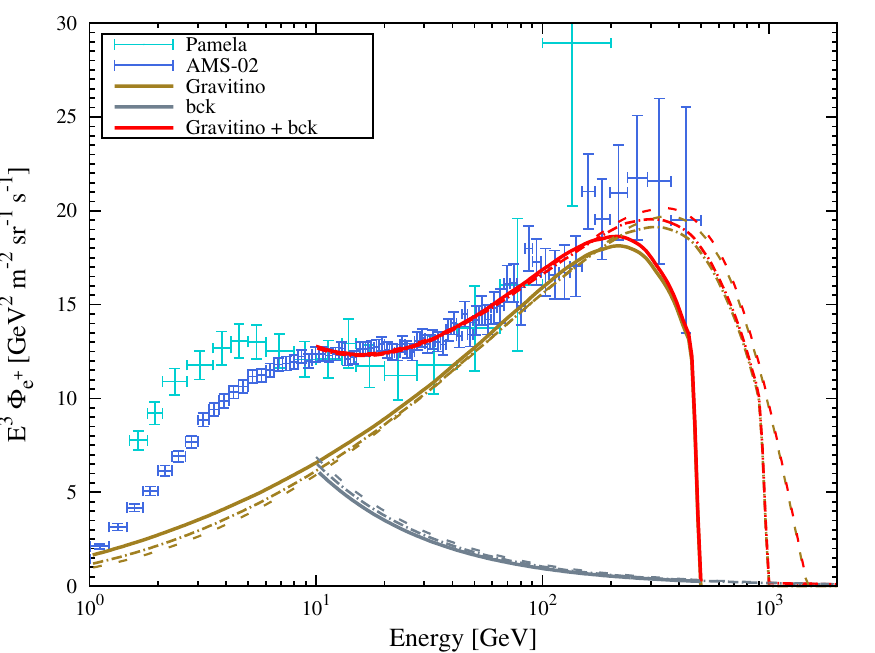}
  }
  \caption{The electron (a) and positron (b) flux measured by AMS-02 and PAMELA (blue and cyan points respectively), and fits using the gravitino decay products + background for three different values of the gravitino mass (in red), $1$ $\TEV$ (continuum line), $2$ $\TEV$ (dot-dashed) and $3$ $\TEV$ (dashed). The background and gravitino contributions are also shown separately in grey and pale brown respectively.}
  \label{fig:inspectra_ep}
\end{figure}

In fig.~\ref{fig:data1} the positron fraction obtained from the best fit values is compared to the PAMELA and AMS-02 data, while a similar comparison for the electron and positron flux are shown in fig.~\ref{fig:inspectra_ep}. Although we do not consider the electron flux measurement in the fit we use the parameters obtained in the fit to compute the electron flux. Notice that a gravitino with $m_{3/2}=1 \TEV$ fails to adjust the highest data point of the electron flux spectrum reported by AMS-02 (see left panel of fig.~\ref{fig:inspectra_ep}), which therefore disfavours scenarios with $m_{3/2}<1\TEV$. From the behaviour of the lines corresponding to $m_{3/2}= 2\TEV$ and $3\TEV$ we can see that we are at the edge of the measured spectra, and that any higher mass value will over predict the data at high energies. 

For completeness we repeated the fit using MIN and MAX propagation models. We find for the three gravitino masses under study that the best fit parameters are compatible with results reported in table \ref{tab:Results-of-AMS}.

%To make this result stronger for fixing values of gravitino mass and lifetime we marginalise the BRs and background parameters. It allow us to find points that provide good enough fit to the data, characterised by a p-value larger than 0.05\footnote{We need to explain this here, BORIS :)}, up to BRs distribution. We find that even using different CR propagation parameters the gravitino lifetime needed to provide an acceptable fit to the AMS-02 data must be shorter than  $\approx 1.5\times10^{26}$ s.

%In the next section  we confront the $\gamma$-ray flux associated to the gravitino parameters suitable to fit AMS-02 data to the EGB recently measured by the Fermi-LAT Collaboration~\cite{Ackecrmann:2014usa}.

\begin{table}
\begin{centering}
\begin{tabular}{|c|c|c|c|c|c|c|}
\hline 
$m_{3/2}$ $[GeV]$ & \multicolumn{2}{c|}{$1000$}  &  \multicolumn{2}{c|}{$2000$} & \multicolumn{2}{c|}{$3000$} \tabularnewline
\hline 
$\chi_{min}^{2}/N$ & \multicolumn{2}{c|}{$0.68$} &  \multicolumn{2}{c|}{$0.62$} & \multicolumn{2}{c|}{$0.61$} \tabularnewline
\hline 
\hline
Free parameter & Best Fit & $\pm1\sigma$ & Best Fit & $\pm1\sigma$ & Best Fit & $\pm1\sigma$ \tabularnewline
\hline
\multirow{2}{*}{$C_{e^{-}}$ $[1/\GEV\, \text{cm}^{2}\, \text{s}\, \text{str}]$} & \multirow{2}{*}{$453.1$} & $511$ & \multirow{2}{*}{$452.9$} & $511$ & \multirow{2}{*}{$452.81$} & $511$\tabularnewline
\cline{3-3} \cline{5-5} \cline{7-7} 
 &  & $400$ &  & $401$ &  & $401$\tabularnewline
\hline 
\multirow{2}{*}{$\gamma_{e^{-}}$} & \multirow{2}{*}{$3.3$} & $3.34$ & \multirow{2}{*}{$3.3$} & $3.34$ & \multirow{2}{*}{$3.3$} & $3.34$\tabularnewline
\cline{3-3} \cline{5-5} \cline{7-7} 
 &  & $3.27$ &  & $3.26$ &  & $3.26$\tabularnewline
\hline 
\multirow{2}{*}{$C_{e^{+}}[1/\GEV\, \text{cm}^{2}\, \text{s}\, \text{str}]$} & \multirow{2}{*}{$40.99$} & $46.65$ & \multirow{2}{*}{$41.36$} & $47.11$ & \multirow{2}{*}{$41.49$} & $47.26$\tabularnewline
\cline{3-3} \cline{5-5} \cline{7-7} 
 &  & $35.88$ &  & $36.22$ &  & $36.4$\tabularnewline
\hline 
\multirow{2}{*}{$\gamma_{e^{+}}$} & \multirow{2}{*}{$3.82$} & $3.89$ & \multirow{2}{*}{$3.8$} & $3.86$ & \multirow{2}{*}{$3.78$} & $3.84$\tabularnewline
\cline{3-3} \cline{5-5} \cline{7-7} 
 &  & $3.78$ &  & $3.74$ &  & $3.73$\tabularnewline
\hline 
\multirow{2}{*}{ $\tau_{3/2}\,[10^{26}\,\text{s}${]} } & \multirow{2}{*}{$0.99$} & $1.04$ & \multirow{2}{*}{$0.79$} & $0.89$ & \multirow{2}{*}{$0.68$} & $0.79$\tabularnewline
\cline{3-3} \cline{5-5} \cline{7-7} 
 &  & $0.78$ &  & $0.66$ &  & $0.59$\tabularnewline
\hline 
\multirow{2}{*}{$BR_{W^{\pm}e^{\mp}}$} & \multirow{2}{*}{0.09} & 0.12 & \multirow{2}{*}{0.05} & 0.11 & \multirow{2}{*}{0.0} & 0.13\tabularnewline
\cline{3-3} \cline{5-5} \cline{7-7} 
 &  & 0.05 &  & 0.0 &  & 0.0\tabularnewline
\hline 
\multirow{2}{*}{$BR_{W^{\pm}\mu^{\mp}}$} & \multirow{2}{*}{0.0} & 0.08 & \multirow{2}{*}{0.0} & 0.19 & \multirow{2}{*}{0.07} & 0.24\tabularnewline
\cline{3-3} \cline{5-5} \cline{7-7} 
 &  & 0.0 &  & 0.0 &  & 0.0\tabularnewline
\hline 
\multirow{2}{*}{$BR_{W^{\pm}\tau^{\mp}}$} & \multirow{2}{*}{0.91} & 0.93 & \multirow{2}{*}{0.82} & 1.0 & \multirow{2}{*}{0.68} & 1.0\tabularnewline
\cline{3-3} \cline{5-5} \cline{7-7} 
 &  & 0.61 &  & 0.44 &  & 0.32\tabularnewline
\hline 
\multirow{2}{*}{$BR_{Z\nu}$} & \multirow{2}{*}{0.0} & 0.29 & \multirow{2}{*}{0.12} & 0.40 & \multirow{2}{*}{0.25} & 0.47\tabularnewline
\cline{3-3} \cline{5-5} \cline{7-7} 
 &  & 0.0 &  & 0.0 &  & 0.0\tabularnewline
\hline 
\multirow{2}{*}{$BR_{H\nu}^{*}$} & \multirow{2}{*}{0.0} & 0.23 & \multirow{2}{*}{0.0} & 0.11 & \multirow{2}{*}{0.0} & 0.13\tabularnewline
\cline{3-3} \cline{5-5} \cline{7-7} 
 &  & 0.0 &  & 0.0 &  & 0.0\tabularnewline
\hline 
\end{tabular}
\par\end{centering}
\caption{Results of AMS fit in a decaying gravitino scenario\label{tab:Results-of-AMS} using MED propagation model. During the fit we have considered $BR_{H\nu}$ as the dependent branching ratio, however in order to compute their confidence intervals we consider $BR_{Z\nu}$ as the dependent one. We use $N=104$ as the effective number of degree of freedom. The corresponding p-values for each scenario are equal to 0.99.}
\end{table}
%\newpage

%%%%%%%%%%%%%%%%%%%%%%%%%%%%%%%%%%%%%%%%%%%%%%%%%%%%%%%%%%%%%%%%%%%%
\section{ The Extragalactic Gamma-ray Background measured by {\it Fermi}-LAT and gravitino dark matter}
\label{sec:Fermi}
%%%%%%%%%%%%%%%%%%%%%%%%%%%%%%%%%%%%%%%%%%%%%%%%%%%%%%%%%%%%%%%%%%%%

\begin{figure}[t]
   \centering
    \includegraphics[width=0.8\linewidth]{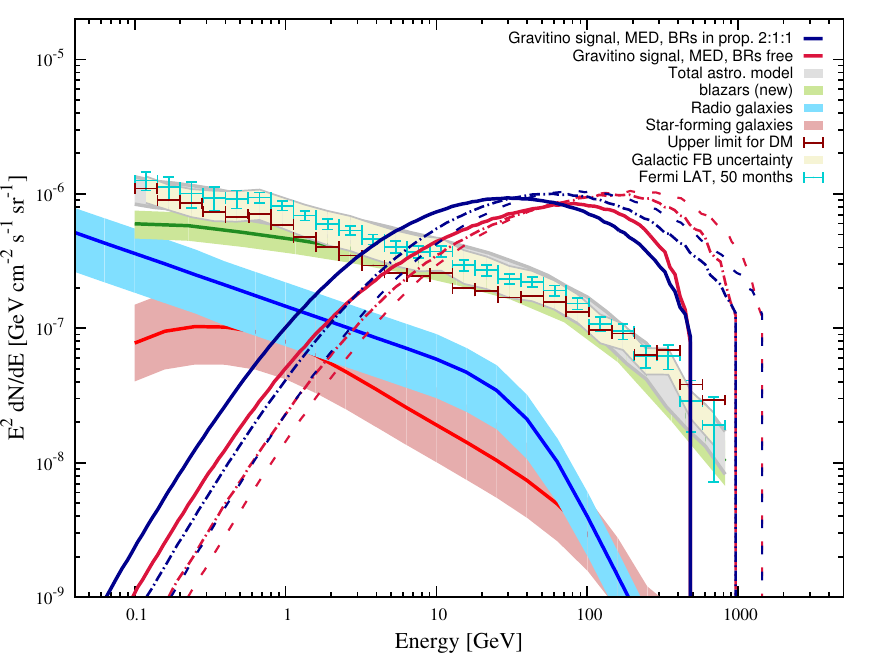}
  \caption{We contrast the predicted $\gamma$-ray flux in the three scenarios obtained from the fit to AMS-02 data to the EGB as measured by {\it Fermi}-LAT, $m_{3/2}=1$ TeV, $\tau_{3/2}=0.99\times 10^{26}$ s; $m_{3/2}=2$ TeV, $\tau_{3/2}=0.78\times10^{26}$ s; and $m_{3/2}=1$ TeV, $\tau_{3/2}=0.67\times10^{26}$ s; solid, dot-dashed, dashed dark-red lines, respectively. The gravitino DM-induced $\gamma$-ray emission clearly exceeds the EGB limits. The EGB as measured by {\it Fermi}-LAT using FG model A (cyan points) and its uncertainty due to Galactic foreground modelling (yellow band). Brown points represent  95\% CL upper limit flux after subtracting from the EGB the average contribution (grey band) from star-forming galaxies (red band~\cite{Ackermann:2012vca}) and radio galaxies (blue band~\cite{2011ApJ...733...66I}) as well as the integrated emission of blazars with EBL absorption as recently modelled in~\cite{AjelloBlazars} (green band).  }
  \label{fig:EGBa}
\end{figure}

The $\gamma$-ray sky has been observed by the {\it Fermi}-LAT Telescope with unprecedented detail. Most of the $\gamma$ rays detected come from: i) point-like or small extended sources and ii) a strong diffuse emission correlated with Galactic structures~\cite{diffuse2}. In addition, a tenuous  diffuse component has been detected, the Isotropic $\gamma$-ray Background (IGRB)~\cite{2010PhRvL.104j1101A}. The origin of the IGRB can be sources that remains below the detection threshold of the {\it Fermi}-LAT, among others, DM decay/annihilation~\cite{Bergstrom:2001jj,Ullio:2002pj} can produce a sizeable contribution\footnote{The main EGB contributors are blazars, star-forming galaxies, diffuse processes as intergalactic shocks~\cite{Colafrancesco:1998us,Loeb:2000na,Zandanel:2013wea}, interactions of ultra high energy CRs with the Extragalactic Background Light (EBL)~\cite{Berezinsky:1975zz}, and CR interactions in small solar system bodies~\cite{2009ApJ...692L..54M}}.  The observed IGRB depends on the point source detection threshold of the instrument. A physical quantity is the total EGB, defined as the combination of resolved sources and the IGRB. The {\it Fermi}-LAT Collaboration has a new determination of the EGB from 50 months of data that expands from 100 MeV to 820 GeV, in fig.~\ref{fig:EGBa} we present this measurement (cyan points with error bars)~\cite{Ackermann:2014usa}\footnote{This is the EGB measurement using FG model A.}. In the extraction of the EGB a relevant source of systematic uncertainty is the modelling of the Galactic foreground (FG) emission\footnote{In order to test this systematic, the fitting procedure to obtain the EGB was applied to many different Galactic FG models.  In particular, reference FG model A is derived from the class of models presented in~\cite{2010PhRvL.104j1101A}. Models B and C have more degrees of freedom in electron and diffusion coefficient, for details on FG models A, B and C we refer to~\cite{Ackermann:2014usa}.}, the yellow band in fig.~\ref{fig:EGBa} represents this uncertainty.

In fig.~\ref{fig:EGBa} in addition to the {\it Fermi}-LAT EGB (cyan points with error bars~\cite{Ackermann:2014usa}) we present the $\gamma$-ray emission due to known EGB contributors, star-forming galaxies (red band~\cite{Ackermann:2012vca}) and radio galaxies (blue band~\cite{2011ApJ...733...66I}) as well as the integrated emission of blazars with EBL absorption as recently modelled in~\cite{AjelloBlazars} (green band). The grey band represents the sum of all these components, it accounts for the observed amplitude and spectral shape of the EGB. From this figure we can see that there is little room for other contributors, such as DM\footnote{Note that Galactic templates of DM decay can be degenerate with ICS templates of the Galactic FG emission model as pointed out in~\cite{cosmowimp}. Nevertheless, we only contrast the gravitino $\gamma$-ray emission to the EGB in an energy range ($>500$ GeV) where the ICS component is not very important, as shown in~\cite{cosmowimp}.}. We compute the 95\% CL upper limit flux subtracting the average emission from non-exotic contributors (black line) to the EGB (cyan points) and adding 1.64 times the sum of data and model uncertainties in quadratures\footnote{We assume Gaussian errors, therefore 95\% of the area of a Gaussian distribution is within 1.64 standard deviations of the mean.}.

%We contrast the $\gamma$-ray flux predicted by the gravitino DM that can fit  the AMS-02 measurements with the latest EGB measurement with the {\it Fermi}-LAT Telescope. The reason to use the EGB is that in the region of interest (RoI) used for subtracting it (the whole sky but the Galactic plane) the uncertainty in the $\gamma$-ray emission associated to Galactic DM decays is lower than in the Galactic plane. It is so because the DM density profiles, for instance NFW and Einasto, are similar and the DM-induced ICS emission is significant smaller than prompt emission. 

%%%%%%%%%%%%%%%%%%%%%%%%%%%%%%
\subsection{$\gamma$-ray flux at Earth from gravitino decays}
%%%%%%%%%%%%%%%%%%%%%%%%%%%%%%

EGB contributions from gravitino DM decays can has 3 different sources: i) the smooth Galactic halo, ii) sub haloes hosted by the Galactic halo and iii) extragalactic structures. The signal from extragalactic gravitino decay is expected to be isotropic, although it is independent on the amount of DM clustering at each given redshift~\cite{2012MNRAS.421L..87S,2002PhR...372....1C,Sefusatti:2014vha}, it is subject of attenuation which increase significantly with the $\gamma$-ray energy~\cite{2012MNRAS.422.3189G}. The emission from the smooth Galactic halo provides a lower limit on the $\gamma$-ray total gravitino contribution to the EGB, we only use this emission in the analysis.   

We calculate the differential flux of $\gamma$ rays from gravitino decays in the Galactic halo by integrating the DM distribution around us along the line of sight. We consider gravitino decays producing $\gamma$ rays in the final state. In this case the flux reads:

\begin{linenomath}
\begin{equation}\label{eq:decayFlux}
 \frac{d\Phi_{\gamma}^{\text{halo}}}{dEd\Omega}=\frac{1}{4\,\pi\,\tau_{3/2}\,m_{3/2}}\,\frac{1}{\Delta\Omega}\,\frac{dN^{\text{total}}_{\gamma}}{dE} \int_{\Delta\Omega}\!\!\cos
 b\,db\,d\ell\int_0^{\infty}\!\! ds\,\rho_{\text{halo}}(r(s,\,b,\,\ell))\,,
\end{equation}
\end{linenomath}

\noindent where $b$ and $\ell$ denote the Galactic latitude and longitude,
respectively, and $s$ denotes the distance from the Solar System. Furthermore, $\Delta \Omega$ is the region of interest (ROI). The radius $r$ in the DM halo density profile of the Milky Way, $\rho_{\text{halo}}$, is expressed in terms of these Galactic coordinates
\begin{linenomath}
\begin{equation}
 r(s,\,b,\,\ell)=\sqrt{s^2+r_{\odot}^2-2\,s\,r_{\odot}\cos{b}\cos{\ell}}\,.
\end{equation}
\end{linenomath}

\noindent For gravitino masses above the Z boson mass, the total number of photons produced in gravitino decays can be expressed as 

\begin{equation}
\frac{dN_{\gamma}^{\text{total}}}{dE}=\sum_{\lambda l_i} BR_{\lambda l_i}\frac{dN_{\lambda l_i\rightarrow \gamma}^{3/2}}{dE},
\label{eq:dndephotona}
\end{equation}

\noindent where $dN_{\lambda l_i\rightarrow \gamma}^{3/2}/dE$ is the photon energy spectrum produced by different gravitino decay channels computed in section \ref{gdecay}, see panel b) of fig.~\ref{fig:inspectra}.

In the next section we contrast the $\gamma$-ray flux predicted by the gravitino DM that can fit  the AMS-02 measurements with the latest EGB measurement with the {\it Fermi}-LAT Telescope.

%In fig.~\ref{fig:EGBa} we contrast the {\it Fermi}-LAT EGB (cyan points with error bars~\cite{Ackermann:2014usa})\footnote{This is the EGB measurement using FG model A.}  to the $\gamma$-ray emission yield by the gravitino DM needed to account for the rise in positron fraction. The gravitino DM-induced $\gamma$-ray flux clearly exceeds the EGB limits. 

\begin{figure}[t]
   \centering
    \includegraphics[width=0.9\linewidth]{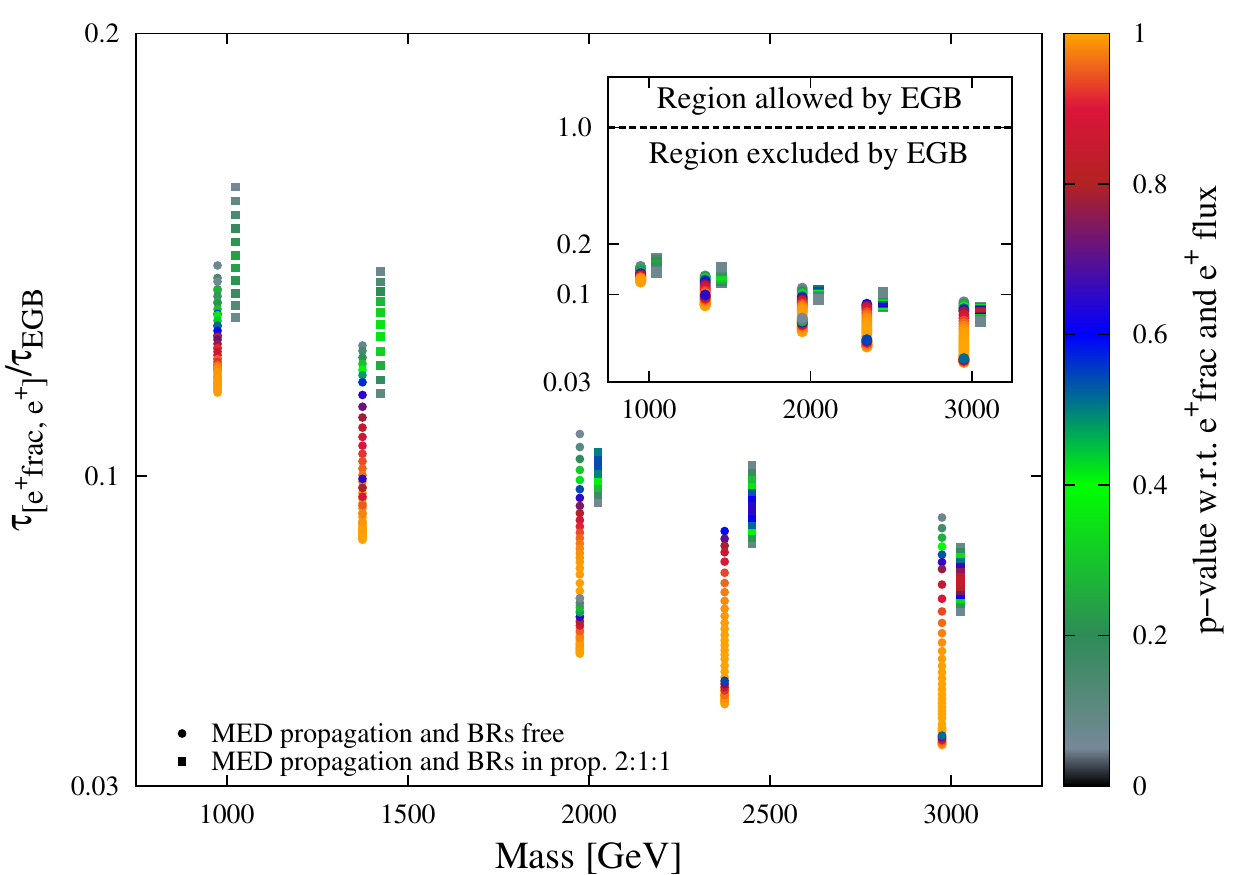}
  \caption{ Comparison between the gravitino lifetimes required to provide acceptable fit (p-value > 0.05, color code) to positron flux and positron fraction $\tau_{[e^+ \text{frac}, e^+]}$ and the lifetimes required to do not overshoot the EGB residual emission, $\tau_{\text{EGB}}$, for different gravitino masses. Points with ratio $\tau_{[e^+ \text{frac}, e^+]}/\tau_{\text{EGB}}$ lower than 1 are excluded since a gravitino with $\tau_{[e^+\text{frac}, e^+]}$ would produce brighter gamma-ray emission than the allowed maximum produced by similar gravitino (same mass and BR distribution) characterised by $\tau_{\text{EGB}}$. Squares represent the points with p-value>0.05 when the restriction 2:1:1 in the BR proportions was considered in the fit, the scenarios with free BRs are represented by circles. All the points reside below the limit  $\tau_{[e^+ \text{frac}, e^+]}/\tau_{\text{EGB}}$=1, are excluded. The smaller plot in the right-upper corner is a zoom out of the bigger plot which points out the size of the exclusion.}
  \label{fig:EGBb}
\end{figure}
%  Limits on gravitino lifetime decay for different BR combinations (channels) preserving the theoretical expectation for gravitinos with mass in the range between 750 and 3000 GeV, $W^{\pm}l^{\mp}_i$:$Z\nu_i$:$H\nu_i$ in proportion 2:1:1~\cite{Ishiwata:2008cu,Grefe:2011dp,Delahaye:2013yqa}. For instance, for the channel ($W^{\pm}\tau^{\mp}$, $Z\nu$, $H\nu$) the proportion 2:1:1 requires the following BR distribution (50\%, 25\%, 25\%). Everything below the solid lines is excluded. The four limits presented aim to enclose a large family of bilinear RpV models. The red band represents a zone where the gravitino DM provides an acceptable explanation of AMS-02 data, characterised by providing fits with p-value larger than 0.05 after marginalising over BRs and background parameters, when using the MED propagation scenario. Light red and pink bands provide good enough explanations to the AMS-02 data as the red band, but the propagation parameters used correspond to MIN and MAX scenarios, respectively. It is clear that any acceptable fit to the AMS-02 data using gravitino DM of bilinear RpV models as unique source of primary electrons and positrons is ruled out by the limits from the Fermi-LAT EGB

%%%%%%%%%%%%%%%%%%%%%%%%%%%%%%%%%%%%%%%%%%%%%%
\section{Contrasting the rise in positron fraction with the EGB through gravitino dark matter}
\label{combination}
%%%%%%%%%%%%%%%%%%%%%%%%%%%%%%%%%%%%%%%%%%%%%%

In fig.~\ref{fig:EGBa} we show the contributions to the EGB from the three gravitino decay scenarios presented in table~\ref{tab:Results-of-AMS}, which provide good fits to the AMS-02 data (gravitino signal red line). We also plot the gravitino-induced gamma-ray emission using the results when we include the 2:1:1 proportion on the BR in the fit (gravitino signal blue line).  They all clearly overshoot the EGB measured by {\it Fermi}-LAT. We already knew that the scenarios with the restriction 2:1:1 on the BRs were excluded because the antiproton limits derived in~\cite{Delahaye:2013yqa}. On the one hand, the results inferred from Fig. 5 show that the EGB limits can be as sensitive as the antiproton measurements in order to test this gravitino scenario. On the other hand, the antiproton limits in~\cite{Delahaye:2013yqa} cannot be directly applied to the gravitino scenarios with free BRs.

As already mentioned, we can see from Fig. 5 that the points giving the best fits to the  AMS-02 data clearly overshot the EGB measurement. Now, in order to be more general, we extend this analysis considering the scenarios that allow a relatively good fit of AMS-02. In this way we check whether we can have a gravitino-induced gamma ray emission consistent with the EGB limits at the cost of a poorer fit to the positron flux and positron fraction measured by AMS-02. In practice, we test points of the parameter space that are not  excluded at 95\% CL by the data, i.e. with a p-value > 0.05. We perform the following analysis to check this out. For fixed gravitino mass and lifetime,  we call this lifetime $\tau_{[e^+ \text{frac}, e^+]}$\footnote{ to point out that comes from an analysis using positron flux and positron fraction.}, we marginalise over the BRs and background parameters, with and without their theoretical prediction 2:1:1, with respect to the positron flux and positron fraction measurements.  We keep the BR distribution corresponding to the largest p-value, then we repeat the procedure for different masses and $\tau_{[e^+ \text{frac}, e^+]}$, in particular we explore 5 gravitino masses, 1.0 TeV, 1.4 TeV, 2.0 TeV, 2.4 TeV, and 3.0 TeV. Next, we use the BR distributions obtained for each of the masses and $\tau_{[e^+ \text{frac}, e^+]}$  to compute the minimum gravitino lifetime allowed by the EGB residual component (brown points in fig.~\ref{fig:EGBa}), $\tau_{\text{EGB}}$. In this way for each single point in the parameter space [mass, $\tau_{[e^+ \text{frac}, e^+]}$, BRs] with p-value > 0.05, we obtain the minimum $\tau_{\text{EGB}}$ which determine if the point [mass, $\tau_{[e^+ \text{frac}, e^+]}$, BRs] is excluded or not, as the ratio $\tau_{[e^+ \text{frac}, e^+]}/\tau_{\text{EGB}}$ is lower or larger than unity, respectively. Results are presented in figure~\ref{fig:EGBb}, where we contrast each of the gravitino DM models that provide an acceptable explanation to the positron rise with the limits on gravitino lifetime from the EGB residual component.  All the points reside below the limit  $\tau_{[e^+ \text{frac}, e^+]}/\tau_{\text{EGB}}$=1, implying that all the gravitino DM scenarios providing relatively good fit to AMS-02 leptonic data are excluded, disregarding their BR distribution or mass.

%The theoretical prediction $W^{\pm}l^{\mp}_i$:$Z\nu_i$:$H\nu_i$ in proportion 2:1:1 requires that the BR of the sum of $W^{\pm}l^{\mp}_i$ channels to be  50\%. From panel b) in fig.~\ref{fig:inspectra} we can see that above 100 GeV the channels that produce less photons are $W^{\pm}e^{\mp}$ and $W^{\pm}\mu^{\mp}$, while $W^{\pm}\tau^{\mp}$ is the dominant. Therefore the tighter constraints would correspond to the BR combination ($W^{\pm}\tau^{\mp}$, $Z\nu$, $H\nu$) = (50\%, 25\%, 25\%) and the softer ones to ($W^{\pm}e^{\mp}$, $Z\nu$, $H\nu$) = (50\%, 25\%, 25\%) and ($W^{\pm}\mu^{\mp}$, $Z\nu$, $H\nu$) = (50\%, 25\%, 25\%), all other possible combinations must fall in between. 

%In fig.~\ref{fig:EGBb} we compare the minimum gravitino lifetime 

%present the limits for these decay channels. These results are in agreement with antiproton constraints within systematic uncertainties in the Galactic propagation model~\cite{Delahaye:2013yqa}, 

Our results are in agreement with the limits for gravitino DM presented in~\cite{Ando:2015qda}. It is worth noting that the limits presented in fig.~\ref{fig:EGBb} are obtained contrasting the gravitino-induced $\gamma$-ray emission from the Galactic smooth halo, excluding the Galactic plane region, to the EGB, for a variety mixture of gravitino decaying channels as required to reproduce AMS-02 leptonic data. This is a different approach to what was used  in~\cite{Ando:2015qda},  where the gravitino-induced gamma-ray emission considered is originated in extragalactic structures and only a single gravitino decay channel is consider $W^{\pm}\mu^{\mp}$. As already mentioned, only considering the Galactic smooth halo contribution provides a minimal contribution to the EGB emission from gravitino decay, but it is enough for the purposes of this work.

%Now, in order to contrast the limits obtained from the EGB with the potential explanations of the positron excess, we show a zone in fig.~\ref{fig:EGBb}  that produce acceptable fits to AMS-02 (red band for MED propagation parameters). The points that belong to this zone are required to have a p-value greater than 0.05, which we assume as a conservative threshold to define an acceptable fit to AMS-02 data\footnote{In other words, any point out of this zone is rejected at 95\% CL as a good explanation of AMS-02 data.}. Notice that for the computation of the p-value we marginalise background parameters and BRs for each point of the parameter space $(m_{3/2},\tau_{3/2})$. Thus, every possible combination of BRs is included in the red band. We also present in fig.~\ref{fig:EGBb} similar zones for MIN (light red), and MAX (pink) propagation scenarios. It is clear that these zones that provide acceptable explanations of AMS-02 data are excluded by the limits from EGB. Therefore we can safely conclude that the gravitino of bilinear RpV SUSY models is excluded as the unique primary source of electrons and positrons needed to explain the rise in the positron fraction.

%Thus, if gravitinos made up the whole DM of the Universe, they can only contribute modestly to the electron and positron fluxes detected at the Earth. 

\section{Conclusions}

Recent AMS-02 results provide new insights on the anomalous rise in positron fraction, presenting the first evidence of a flattening spectrum at higher energies~\cite{AMS}. In addition, the measurement of electron and positron fluxes allow a better characterisation of the phenomenon behind the positron fraction rise~\cite{AMS1}. We model positron and electron fluxes as decreasing-with-energy power laws plus an extra source injecting equally electrons and positrons in the ISM, we test the gravitino of RpV SUSY models as the source term (e.g.,~\cite{Buchmuller:2007ui,Grefe:2008zz,Choi:2010jt,Restrepo:2011rj,Diaz:2011pc,Cottin:2014cca}). We compute the electron, positron and $\gamma$-ray spectra produced by gravitino DM decays as would be detected in the Earth. We have shown that the gravitino DM with mass in the range $1-3$ TeV and lifetime of $\sim 1 - 0.7 \times 10^{26}$s, mainly decaying to $W^{\pm}\tau^{\mp}$ can reproduce the energy spectra of the positron fraction and the electron and positron fluxes at the same time. In order to test the viability of these scenarios we study the corresponding $\gamma$-ray emission and we find that these scenarios clearly exceed the observed EGB. Then, we use the most recent {\it Fermi}-LAT EGB measurement~\cite{Ackermann:2014usa} and a model of its known contributors, star-forming galaxies, radio galaxies, and blazars\footnote{The integrated emission of blazars include the EBL absorption we use the model in~\cite{AjelloBlazars}} to compare with the emission from every single gravitino DM able to provide an acceptable explanation of the AMS-02 leptonic measurements, characterised by their p-value larger than 0.05. We find all the points overshoot the EGB limits by about one order of magnitud, this excludes the possibility of gravitino DM in bilinear RpV models as unique source of primary positrons\footnote{Note that we are not including the gravitino two body decay into heavy higgs since, in general, they are heavier than the LSP. Also notice that this conclusion depends strongly on the modelling of the electron and positron background, which is not well understood yet.}. Thus, if those gravitinos made up the whole DM of the Universe, they can only contribute modestly to the highest energetic electron and positron fluxes detected at the Earth.

%The corresponding $\gamma$-ray emission significantly exceeds most recent {\it Fermi}-LAT EGB measurement~\cite{Ackermann:2014usa} using 50 months of data, excluding in this way the possibility of gravitino DM in bilinear RpV models as unique source of primary positrons\footnote{Note that we are not including the gravitino two body decay into heavy higgs since, in general, they are heavier than the LSP.}. We use the EGB and a model of its known contributors, star-forming galaxies, radio galaxies, and blazars\footnote{The integrated emission of blazars include the EBL absorption we use the model in~\cite{AjelloBlazars}}, to restrict the gravitino lifetime. Thus, if gravitinos of bilinear RpV models made up the whole DM of the Universe, they can only contribute modestly to the highest energetic electron and positron fluxes detected at the Earth. 

\section*{Acknowledgments}

{\small 
The authors are thankful to Andrea Albert, Borut Bajc, Marco Cirelli, Michael Grefe, Luis Labarga, Carlos Muñoz, Paolo Panci, Frank Steffen, and Gabriela Zaharijas for useful comments, and Marco Ajello for providing the EGB model contributions of figure~\ref{fig:EGBb}. Also we thank the anonymous referee of PDMU for recommendations in the fitting procedure.
 This work was supported by Conicyt Anillo grant ACT1102. GAGV thanks for the support of the Spanish MINECO's Consolider-Ingenio 2010 Programme under grant MultiDark CSD2009-00064 also the partial support by MINECO under grant FPA2012-34694. BP also thanks for the support of the State of S\~{a}o Paulo Research Foundation (FAPESP). The work of NV was supported by CONICYT FONDECYT/POSTDOCTORADO/3140559.
}
%%%%%%%%%%%%%%%%%%%%%%%%%%%%%%%%%%%%%%%%%%%%%%%%%%%%%%%%%%%%%%%%%%

\bibliographystyle{JHEP}
\bibliography{Gravitino_AMS_Fermi_arxiv_rep_german_v1_intro_revisited}

\providecommand{\href}[2]{#2}\begingroup\raggedright\begin{thebibliography}{10}

\bibitem{2012JMPh....3.1152R}
M.~{Roos}, {\it {Astrophysical and Cosmological Probes of Dark Matter}},  {\em
  Journal of Modern Physics} {\bf 3} (2012) 1152--1171,
  [\href{http://xxx.lanl.gov/abs/1208.3662}{{\tt arXiv:1208.3662}}].

\bibitem{Bertone09}
G.~{Bertone}, D.~{Hooper}, and J.~{Silk}, {\it {Particle dark matter: evidence,
  candidates and constraints}},  {\em \physrep} {\bf 405} (Jan., 2005)
  279--390, [\href{http://xxx.lanl.gov/abs/hep-ph/0404175}{{\tt
  hep-ph/0404175}}].

\bibitem{Munoz:2003gx}
C.~Munoz, {\it {Dark matter detection in the light of recent experimental
  results}},  {\em Int.J.Mod.Phys.} {\bf A19} (2004) 3093--3170,
  [\href{http://xxx.lanl.gov/abs/hep-ph/0309346}{{\tt hep-ph/0309346}}].

\bibitem{Adriani:2008zr}
{\bf PAMELA} Collaboration, O.~Adriani et~al., {\it {An anomalous positron
  abundance in cosmic rays with energies 1.5-100 GeV}},  {\em Nature} {\bf 458}
  (2009) 607--609, [\href{http://xxx.lanl.gov/abs/0810.4995}{{\tt
  arXiv:0810.4995}}].

\bibitem{FermiPositron}
{\bf The Fermi LAT} Collaboration, M.~Ackermann et~al., {\it {Measurement of
  Separate Cosmic-Ray Electron and Positron Spectra with the Fermi Large Area
  Telescope}},  {\em Phys.Rev.Lett.} {\bf 108} (Jan., 2012) 011103,
  [\href{http://xxx.lanl.gov/abs/1109.0521}{{\tt arXiv:1109.0521}}].

\bibitem{Aguilar:2013qda}
{\bf AMS Collaboration} Collaboration, M.~Aguilar et~al., {\it {First Result
  from the Alpha Magnetic Spectrometer on the International Space Station:
  Precision Measurement of the Positron Fraction in Primary Cosmic Rays of
  0.5-350 GeV}},  {\em Phys.Rev.Lett.} {\bf 110} (2013) 141102.

\bibitem{Feng:2013zca}
L.~Feng, R.-Z. Yang, H.-N. He, T.-K. Dong, Y.-Z. Fan, et~al., {\it {AMS-02
  positron excess: new bounds on dark matter models and hint for primary
  electron spectrum hardening}},  {\em Phys.Lett.} {\bf B728} (2014) 250--255,
  [\href{http://xxx.lanl.gov/abs/1303.0530}{{\tt arXiv:1303.0530}}].

\bibitem{Bergstrom:2013jra}
L.~Bergstrom, T.~Bringmann, I.~Cholis, D.~Hooper, and C.~Weniger, {\it {New
  limits on dark matter annihilation from AMS cosmic ray positron data}},  {\em
  Phys.Rev.Lett.} {\bf 111} (2013) 171101,
  [\href{http://xxx.lanl.gov/abs/1306.3983}{{\tt arXiv:1306.3983}}].

\bibitem{Ibarra:2013cra}
A.~Ibarra, D.~Tran, and C.~Weniger, {\it {Indirect Searches for Decaying Dark
  Matter}},  {\em Int.J.Mod.Phys.} {\bf A28} (2013) 1330040,
  [\href{http://xxx.lanl.gov/abs/1307.6434}{{\tt arXiv:1307.6434}}].

\bibitem{Choi:2013oaa}
K.-Y. Choi, B.~Kyae, and C.~S. Shin, {\it {Decaying WIMP dark matter for AMS-02
  cosmic positron excess}},  {\em Phys.Rev.} {\bf D89} (2014), no.~5 055002,
  [\href{http://xxx.lanl.gov/abs/1307.6568}{{\tt arXiv:1307.6568}}].

\bibitem{Hryczuk:2014hpa}
A.~Hryczuk, I.~Cholis, R.~Iengo, M.~Tavakoli, and P.~Ullio, {\it {Indirect
  Detection Analysis: Wino Dark Matter Case Study}},  {\em JCAP} {\bf 1407}
  (2014) 031, [\href{http://xxx.lanl.gov/abs/1401.6212}{{\tt
  arXiv:1401.6212}}].

\bibitem{Baek:2014awa}
S.~Baek, H.~Okada, and T.~Toma, {\it {Radiative lepton model and dark matter
  with global $U(1)'$ symmetry}},  {\em Phys.Lett.} {\bf B732} (2014) 85--90,
  [\href{http://xxx.lanl.gov/abs/1401.6921}{{\tt arXiv:1401.6921}}].

\bibitem{Zhao:2014nsa}
Y.~Zhao and K.~M. Zurek, {\it {Indirect Detection Signatures for the Origin of
  Asymmetric Dark Matter}},  {\em JHEP} {\bf 1407} (2014) 017,
  [\href{http://xxx.lanl.gov/abs/1401.7664}{{\tt arXiv:1401.7664}}].

\bibitem{Cirelli:2008pk}
M.~Cirelli, M.~Kadastik, M.~Raidal, and A.~Strumia, {\it {Model-independent
  implications of the e+-, anti-proton cosmic ray spectra on properties of Dark
  Matter}},  {\em Nucl.Phys.} {\bf B813} (2009) 1--21,
  [\href{http://xxx.lanl.gov/abs/0809.2409}{{\tt arXiv:0809.2409}}].

\bibitem{Ibarra:2013zia}
A.~Ibarra, A.~S. Lamperstorfer, and J.~Silk, {\it {Dark matter annihilations
  and decays after the AMS-02 positron measurements}},  {\em Phys.Rev.} {\bf
  D89} (2014) 063539, [\href{http://xxx.lanl.gov/abs/1309.2570}{{\tt
  arXiv:1309.2570}}].

\bibitem{Dev:2013hka}
P.~S.~B. Dev, D.~K. Ghosh, N.~Okada, and I.~Saha, {\it {Neutrino Mass and Dark
  Matter in light of recent AMS-02 results}},  {\em Phys.Rev.} {\bf D89} (2014)
  095001, [\href{http://xxx.lanl.gov/abs/1307.6204}{{\tt arXiv:1307.6204}}].

\bibitem{Ibe:2013jya}
M.~Ibe, S.~Matsumoto, S.~Shirai, and T.~T. Yanagida, {\it {AMS-02 Positrons
  from Decaying Wino in the Pure Gravity Mediation Model}},  {\em JHEP} {\bf
  1307} (2013) 063, [\href{http://xxx.lanl.gov/abs/1305.0084}{{\tt
  arXiv:1305.0084}}].

\bibitem{Baek:2014goa}
S.~Baek, P.~Ko, W.-I. Park, and Y.~Tang, {\it {Indirect and direct signatures
  of Higgs portal decaying vector dark matter for positron excess in cosmic
  rays}},  {\em JCAP} {\bf 1406} (2014) 046,
  [\href{http://xxx.lanl.gov/abs/1402.2115}{{\tt arXiv:1402.2115}}].

\bibitem{ArkaniHamed:2008qn}
N.~Arkani-Hamed, D.~P. Finkbeiner, T.~R. Slatyer, and N.~Weiner, {\it {A Theory
  of Dark Matter}},  {\em Phys.Rev.} {\bf D79} (2009) 015014,
  [\href{http://xxx.lanl.gov/abs/0810.0713}{{\tt arXiv:0810.0713}}].

\bibitem{Ibarra:2009dr}
A.~Ibarra, D.~Tran, and C.~Weniger, {\it {Decaying Dark Matter in Light of the
  PAMELA and Fermi LAT Data}},  {\em JCAP} {\bf 1001} (2010) 009,
  [\href{http://xxx.lanl.gov/abs/0906.1571}{{\tt arXiv:0906.1571}}].

\bibitem{Ibe:2013nka}
M.~Ibe, S.~Iwamoto, S.~Matsumoto, T.~Moroi, and N.~Yokozaki, {\it {Recent
  Result of the AMS-02 Experiment and Decaying Gravitino Dark Matter in Gauge
  Mediation}},  {\em JHEP} {\bf 1308} (2013) 029,
  [\href{http://xxx.lanl.gov/abs/1304.1483}{{\tt arXiv:1304.1483}}].

\bibitem{Delahaye:2013yqa}
T.~Delahaye and M.~Grefe, {\it {Antiproton limits on decaying gravitino dark
  matter}},  {\em JCAP} {\bf 1312} (2013) 045,
  [\href{http://xxx.lanl.gov/abs/1305.7183}{{\tt arXiv:1305.7183}}].

\bibitem{2012PhRvD..86h3506C}
M.~{Cirelli}, E.~{Moulin}, P.~{Panci}, P.~D. {Serpico}, and A.~{Viana}, {\it
  {Gamma ray constraints on decaying dark matter}},  {\em \prd} {\bf 86} (Oct.,
  2012) 083506, [\href{http://xxx.lanl.gov/abs/1205.5283}{{\tt
  arXiv:1205.5283}}].

\bibitem{Ando:2015qda}
S.~Ando and K.~Ishiwata, {\it {Constraints on decaying dark matter from the
  extragalactic gamma-ray background}},  {\em JCAP} {\bf 1505} (2015), no.~05
  024, [\href{http://xxx.lanl.gov/abs/1502.0200}{{\tt arXiv:1502.0200}}].

\bibitem{2014arXiv1403.6111P}
M.~A. {Perez-Garcia} and J.~{Silk}, {\it {Constraining decaying dark matter
  with neutron stars}},  {\em ArXiv e-prints} (Mar., 2014)
  [\href{http://xxx.lanl.gov/abs/1403.6111}{{\tt arXiv:1403.6111}}].

\bibitem{Grefe:2011dp}
M.~Grefe, {\it {Unstable Gravitino Dark Matter - Prospects for Indirect and
  Direct Detection}},  \href{http://xxx.lanl.gov/abs/1111.6779}{{\tt
  arXiv:1111.6779}}.

\bibitem{Grefe:2015jva}
T.~Delahaye and M.~Grefe, {\it {Antideuterons from Decaying Gravitino Dark
  Matter}},  {\em JCAP} {\bf 1507} (2015), no.~07 012,
  [\href{http://xxx.lanl.gov/abs/1503.0110}{{\tt arXiv:1503.0110}}].

\bibitem{Dal:2014nda}
L.~A. Dal and A.~R. Raklev, {\it {Antideuteron Limits on Decaying Dark Matter
  with a Tuned Formation Model}},  {\em Phys. Rev.} {\bf D89} (2014), no.~10
  103504, [\href{http://xxx.lanl.gov/abs/1402.6259}{{\tt arXiv:1402.6259}}].

\bibitem{Monteux:2014tia}
A.~Monteux, E.~Carlson, and J.~Cornell, {\it {Gravitino Dark Matter and Flavor
  Symmetries}},  {\em JHEP} {\bf 08} (2014) 047,
  [\href{http://xxx.lanl.gov/abs/1404.5952}{{\tt arXiv:1404.5952}}].

\bibitem{DiMauro:2014iia}
M.~Di~Mauro, F.~Donato, N.~Fornengo, R.~Lineros, and A.~Vittino, {\it
  {Interpretation of AMS-02 electrons and positrons data}},  {\em JCAP} {\bf
  1404} (2014) 006, [\href{http://xxx.lanl.gov/abs/1402.0321}{{\tt
  arXiv:1402.0321}}].

\bibitem{Hooper:2008kg}
D.~Hooper, P.~Blasi, and P.~D. Serpico, {\it {Pulsars as the Sources of High
  Energy Cosmic Ray Positrons}},  {\em JCAP} {\bf 0901} (2009) 025,
  [\href{http://xxx.lanl.gov/abs/0810.1527}{{\tt arXiv:0810.1527}}].

\bibitem{Igor}
I.~V. {Moskalenko}, {\it {Cosmic Rays in the Milky Way and Beyond}},  {\em
  Nuclear Physics B Proceedings Supplements} {\bf 243} (Oct., 2013) 85--91,
  [\href{http://xxx.lanl.gov/abs/1308.5482}{{\tt arXiv:1308.5482}}].

\bibitem{AMS}
{\bf AMS} Collaboration, L.~Accardo et~al., {\it High statistics measurement of
  the positron fraction in primary cosmic rays of 0.5\char21{}500 gev with the
  alpha magnetic spectrometer on the international space station},  {\em Phys.
  Rev. Lett.} {\bf 113} (Sep, 2014) 121101.

\bibitem{AMS1}
{\bf AMS} Collaboration, M.~Aguilar et~al., {\it Electron and positron fluxes
  in primary cosmic rays measured with the alpha magnetic spectrometer on the
  international space station},  {\em Phys. Rev. Lett.} {\bf 113} (Sep, 2014)
  121102.

\bibitem{Takayama:2000uz}
F.~Takayama and M.~Yamaguchi, {\it {Gravitino dark matter without R-parity}},
  {\em Phys.Lett.} {\bf B485} (2000) 388--392,
  [\href{http://xxx.lanl.gov/abs/hep-ph/0005214}{{\tt hep-ph/0005214}}].

\bibitem{Buchmuller:2007ui}
W.~Buchmuller, L.~Covi, K.~Hamaguchi, A.~Ibarra, and T.~Yanagida, {\it
  {Gravitino Dark Matter in R-Parity Breaking Vacua}},  {\em JHEP} {\bf 0703}
  (2007) 037, [\href{http://xxx.lanl.gov/abs/hep-ph/0702184}{{\tt
  hep-ph/0702184}}].

\bibitem{Grefe:2008zz}
M.~Grefe, {\it {Neutrino signals from gravitino dark matter with broken
  R-parity}},  \href{http://xxx.lanl.gov/abs/1111.6041}{{\tt arXiv:1111.6041}}.

\bibitem{Choi:2010jt}
K.-Y. Choi, D.~Restrepo, C.~E. Yaguna, and O.~Zapata, {\it {Indirect detection
  of gravitino dark matter including its three-body decays}},  {\em JCAP} {\bf
  1010} (2010) 033, [\href{http://xxx.lanl.gov/abs/1007.1728}{{\tt
  arXiv:1007.1728}}].

\bibitem{Restrepo:2011rj}
D.~Restrepo, M.~Taoso, J.~Valle, and O.~Zapata, {\it {Gravitino dark matter and
  neutrino masses with bilinear R-parity violation}},  {\em Phys.Rev.} {\bf
  D85} (2012) 023523, [\href{http://xxx.lanl.gov/abs/1109.0512}{{\tt
  arXiv:1109.0512}}].

\bibitem{Diaz:2011pc}
M.~A. Diaz, S.~G. Saenz, and B.~Koch, {\it {Gravitino Dark Matter and Neutrino
  Masses in Partial Split Supersymmetry}},  {\em Phys.Rev.} {\bf D84} (2011)
  055007, [\href{http://xxx.lanl.gov/abs/1106.0308}{{\tt arXiv:1106.0308}}].

\bibitem{Cottin:2014cca}
G.~Cottin, M.~A. Diaz, M.~J. Guzman, and B.~Panes, {\it {Gravitino Dark Matter
  in Split Supersymmetry with Bilinear R-Parity Violation}},  {\em Eur. Phys.
  J.} {\bf C74} (2014), no.~11 3138,
  [\href{http://xxx.lanl.gov/abs/1406.2368}{{\tt arXiv:1406.2368}}].

\bibitem{Bomark:2009zm}
N.-E. Bomark, S.~Lola, P.~Osland, and A.~Raklev, {\it {Photon, Neutrino and
  Charged Particle Spectra from R-violating Gravitino Decays}},  {\em
  Phys.Lett.} {\bf B686} (2010) 152--161,
  [\href{http://xxx.lanl.gov/abs/0911.3376}{{\tt arXiv:0911.3376}}].

\bibitem{Bajc:2010qj}
B.~Bajc, T.~Enkhbat, D.~K. Ghosh, G.~Senjanovic, and Y.~Zhang, {\it {MSSM in
  view of PAMELA and Fermi-LAT}},  {\em JHEP} {\bf 1005} (2010) 048,
  [\href{http://xxx.lanl.gov/abs/1002.3631}{{\tt arXiv:1002.3631}}].

\bibitem{Fayet:1981sq}
P.~Fayet, {\it Experimental consequences of supersymmetry},  in {\em
  Proceedings of the 16th Rencontre de Moriond} (J.~{Tran Thanh Van}, ed.),
  vol.~1, pp.~347--367, Editions Frontieres, 1981.

\bibitem{Giudice:1999am}
G.~Giudice, A.~Riotto, and I.~Tkachev, {\it {Thermal and nonthermal production
  of gravitinos in the early universe}},  {\em JHEP} {\bf 9911} (1999) 036,
  [\href{http://xxx.lanl.gov/abs/hep-ph/9911302}{{\tt hep-ph/9911302}}].

\bibitem{Bolz:2000fu}
M.~Bolz, A.~Brandenburg, and W.~Buchmuller, {\it {Thermal production of
  gravitinos}},  {\em Nucl. Phys.} {\bf B606} (2001) 518--544,
  [\href{http://xxx.lanl.gov/abs/hep-ph/0012052}{{\tt hep-ph/0012052}}].

\bibitem{Pradler:2006qh}
J.~Pradler and F.~D. Steffen, {\it {Thermal gravitino production and collider
  tests of leptogenesis}},  {\em Phys.Rev.} {\bf D75} (2007) 023509,
  [\href{http://xxx.lanl.gov/abs/hep-ph/0608344}{{\tt hep-ph/0608344}}].

\bibitem{Rychkov:2007uq}
V.~S. Rychkov and A.~Strumia, {\it {Thermal production of gravitinos}},  {\em
  Phys.Rev.} {\bf D75} (2007) 075011,
  [\href{http://xxx.lanl.gov/abs/hep-ph/0701104}{{\tt hep-ph/0701104}}].

\bibitem{Moreau:2001sr}
G.~Moreau and M.~Chemtob, {\it {R-parity violation and the cosmological
  gravitino problem}},  {\em Phys.Rev.} {\bf D65} (2002) 024033,
  [\href{http://xxx.lanl.gov/abs/hep-ph/0107286}{{\tt hep-ph/0107286}}].

\bibitem{Ishiwata:2008cu}
K.~Ishiwata, S.~Matsumoto, and T.~Moroi, {\it {High Energy Cosmic Rays from the
  Decay of Gravitino Dark Matter}},  {\em Phys.Rev.} {\bf D78} (2008) 063505,
  [\href{http://xxx.lanl.gov/abs/0805.1133}{{\tt arXiv:0805.1133}}].

\bibitem{PYTHIA8}
{Sj\"ostrand, T. and Mrenna, S. and Skands, P.}, {\it {A Brief Introduction to
  PYTHIA 8.1}},  {\em Comput. Phys. Commun.} {\bf 178} (2008) 852.

\bibitem{Ackermann:2014usa}
{\bf Fermi-LAT} Collaboration, M.~Ackermann et~al., {\it {The spectrum of
  isotropic diffuse gamma-ray emission between 100 MeV and 820 GeV}},  {\em
  Astrophys. J.} {\bf 799} (2015) 86,
  [\href{http://xxx.lanl.gov/abs/1410.3696}{{\tt arXiv:1410.3696}}].

\bibitem{AjelloBlazars}
M.~Ajello et~al., {\it {The Origin of the Extragalactic Gamma-Ray Background
  and Implications for Dark-Matter Annihilation}},  {\em Astrophys. J.} {\bf
  800} (2015), no.~2 L27, [\href{http://xxx.lanl.gov/abs/1501.05301}{{\tt
  arXiv:1501.05301}}].

\bibitem{Ishiwata:2009vx}
K.~Ishiwata, S.~Matsumoto, and T.~Moroi, {\it {High Energy Cosmic Rays from
  Decaying Supersymmetric Dark Matter}},  {\em JHEP} {\bf 0905} (2009) 110,
  [\href{http://xxx.lanl.gov/abs/0903.0242}{{\tt arXiv:0903.0242}}].

\bibitem{Hirsch:2000ef}
M.~Hirsch, M.~Diaz, W.~Porod, J.~Romao, and J.~Valle, {\it {Neutrino masses and
  mixings from supersymmetry with bilinear R parity violation: A Theory for
  solar and atmospheric neutrino oscillations}},  {\em Phys.Rev.} {\bf D62}
  (2000) 113008, [\href{http://xxx.lanl.gov/abs/hep-ph/0004115}{{\tt
  hep-ph/0004115}}].

\bibitem{Strong:2007nh}
A.~W. Strong, I.~V. Moskalenko, and V.~S. Ptuskin, {\it {Cosmic-ray propagation
  and interactions in the Galaxy}},  {\em Ann.Rev.Nucl.Part.Sci.} {\bf 57}
  (2007) 285--327, [\href{http://xxx.lanl.gov/abs/astro-ph/0701517}{{\tt
  astro-ph/0701517}}].

\bibitem{cirelli}
M.~{Cirelli}, G.~{Corcella}, A.~{Hektor}, G.~{H{\"u}tsi}, M.~{Kadastik},
  P.~{Panci}, M.~{Raidal}, F.~{Sala}, and A.~{Strumia}, {\it {PPPC 4 DM ID: a
  poor particle physicist cookbook for dark matter indirect detection}},  {\em
  \jcap} {\bf 3} (Mar., 2011) 51,
  [\href{http://xxx.lanl.gov/abs/1012.4515}{{\tt arXiv:1012.4515}}].

\bibitem{Delh2008}
T.~{Delahaye}, R.~{Lineros}, F.~{Donato}, N.~{Fornengo}, and P.~{Salati}, {\it
  {Positrons from dark matter annihilation in the galactic halo: Theoretical
  uncertainties}},  {\em \prd} {\bf 77} (Mar., 2008) 063527,
  [\href{http://xxx.lanl.gov/abs/0712.2312}{{\tt arXiv:0712.2312}}].

\bibitem{nfw}
J.~F. {Navarro}, C.~S. {Frenk}, and S.~D.~M. {White}, {\it {The Structure of
  Cold Dark Matter Halos}},  {\em \apj} {\bf 462} (May, 1996) 563,
  [\href{http://xxx.lanl.gov/abs/astro-ph/9508025}{{\tt astro-ph/9508025}}].

\bibitem{Gillessen09}
S.~{Gillessen}, F.~{Eisenhauer}, S.~{Trippe}, T.~{Alexander}, R.~{Genzel},
  F.~{Martins}, and T.~{Ott}, {\it {Monitoring Stellar Orbits Around the
  Massive Black Hole in the Galactic Center}},  {\em \apj} {\bf 692} (Feb.,
  2009) 1075--1109, [\href{http://xxx.lanl.gov/abs/0810.4674}{{\tt
  arXiv:0810.4674}}].

\bibitem{Lavalle:2014kca}
J.~Lavalle, D.~Maurin, and A.~Putze, {\it {Direct constraints on diffusion
  models from cosmic-ray positron data: Excluding the minimal model for dark
  matter searches}},  {\em Phys. Rev.} {\bf D90} (2014) 081301,
  [\href{http://xxx.lanl.gov/abs/1407.2540}{{\tt arXiv:1407.2540}}].

\bibitem{Giesen:2015ufa}
G.~Giesen, M.~Boudaud, Y.~Génolini, V.~Poulin, M.~Cirelli, P.~Salati, and
  P.~D. Serpico, {\it {AMS-02 antiprotons, at last! Secondary astrophysical
  component and immediate implications for Dark Matter}},  {\em JCAP} {\bf
  1509} (2015), no.~09 023, [\href{http://xxx.lanl.gov/abs/1504.04276}{{\tt
  arXiv:1504.04276}}].

\bibitem{Battiston20146}
R.~Battiston, {\it Precision measurements of in cosmic ray with the alpha
  magnetic spectrometer on the iss},  {\em Physics of the Dark Universe} {\bf
  4} (2014).

\bibitem{LLyons1986}
L.~Lyons, {\em Statistics for nuclear and particle physicists}.
\newblock Cambridge University Press, 1986.

\bibitem{2009PhRvL.102r1101A}
{\bf The Fermi LAT} Collaboration, A.~A. {Abdo} et~al., {\it {Measurement of
  the Cosmic Ray e$^{+}$+e$^{-}$ Spectrum from 20GeV to 1TeV with the Fermi
  Large Area Telescope}},  {\em Physical Review Letters} {\bf 102} (May, 2009)
  181101, [\href{http://xxx.lanl.gov/abs/0905.0025}{{\tt arXiv:0905.0025}}].

\bibitem{2011PhRvL.106t1101A}
{\bf PAMELA} Collaboration, O.~{Adriani} et~al., {\it {Cosmic-Ray Electron Flux
  Measured by the PAMELA Experiment between 1 and 625 GeV}},  {\em Physical
  Review Letters} {\bf 106} (May, 2011) 201101,
  [\href{http://xxx.lanl.gov/abs/1103.2880}{{\tt arXiv:1103.2880}}].

\bibitem{diffuse2}
{\bf The Fermi LAT} Collaboration, M.~Ackermann et~al., {\it {Fermi-LAT
  Observations of the Diffuse {$\gamma$}-Ray Emission: Implications for Cosmic
  Rays and the Interstellar Medium}},  {\em \apj} {\bf 750} (May, 2012) 3,
  [\href{http://xxx.lanl.gov/abs/1202.4039}{{\tt arXiv:1202.4039}}].

\bibitem{2010PhRvL.104j1101A}
{\bf The Fermi LAT} Collaboration, M.~Ackermann et~al., {\it {Spectrum of the
  Isotropic Diffuse Gamma-Ray Emission Derived from First-Year Fermi Large Area
  Telescope Data}},  {\em Physical Review Letters} {\bf 104} (Mar., 2010)
  101101, [\href{http://xxx.lanl.gov/abs/1002.3603}{{\tt arXiv:1002.3603}}].

\bibitem{Bergstrom:2001jj}
L.~Bergstrom, J.~Edsjo, and P.~Ullio, {\it {Spectral gamma-ray signatures of
  cosmological dark matter annihilation}},  {\em Phys.Rev.Lett.} {\bf 87}
  (2001) 251301, [\href{http://xxx.lanl.gov/abs/astro-ph/0105048}{{\tt
  astro-ph/0105048}}].

\bibitem{Ullio:2002pj}
P.~Ullio, L.~Bergstrom, J.~Edsjo, and C.~G. Lacey, {\it {Cosmological dark
  matter annihilations into gamma-rays - a closer look}},  {\em Phys.Rev.} {\bf
  D66} (2002) 123502, [\href{http://xxx.lanl.gov/abs/astro-ph/0207125}{{\tt
  astro-ph/0207125}}].

\bibitem{Colafrancesco:1998us}
S.~Colafrancesco and P.~Blasi, {\it {Clusters of galaxies and the diffuse
  gamma-ray background}},  {\em Astropart.Phys.} {\bf 9} (1998) 227--246,
  [\href{http://xxx.lanl.gov/abs/astro-ph/9804262}{{\tt astro-ph/9804262}}].

\bibitem{Loeb:2000na}
A.~Loeb and E.~Waxman, {\it {Gamma-ray background from structure formation in
  the intergalactic medium}},  {\em Nature} {\bf 405} (2000) 156,
  [\href{http://xxx.lanl.gov/abs/astro-ph/0003447}{{\tt astro-ph/0003447}}].

\bibitem{Zandanel:2013wea}
F.~Zandanel and S.~Ando, {\it {Constraints on diffuse gamma-ray emission from
  structure formation processes in the Coma cluster}},  {\em
  Mon.Not.Roy.Astron.Soc.} {\bf 440} (2014) 663--671,
  [\href{http://xxx.lanl.gov/abs/1312.1493}{{\tt arXiv:1312.1493}}].

\bibitem{Berezinsky:1975zz}
V.~Berezinsky and A.~Y. Smirnov, {\it {Cosmic neutrinos of ultra-high energies
  and detection possibility}},  {\em Astrophys.Space Sci.} {\bf 32} (1975)
  461--482.

\bibitem{2009ApJ...692L..54M}
I.~V. {Moskalenko} and T.~A. {Porter}, {\it {Isotropic Gamma-Ray Background:
  Cosmic-Ray-Induced Albedo from Debris in the Solar System?}},  {\em \apjl}
  {\bf 692} (Feb., 2009) L54--L57,
  [\href{http://xxx.lanl.gov/abs/0901.0304}{{\tt arXiv:0901.0304}}].

\bibitem{Ackermann:2012vca}
{\bf Fermi LAT} Collaboration, M.~Ackermann et~al., {\it {GeV Observations of
  Star-forming Galaxies with \textit{Fermi} LAT}},  {\em Astrophys.J.} {\bf
  755} (2012) 164, [\href{http://xxx.lanl.gov/abs/1206.1346}{{\tt
  arXiv:1206.1346}}].

\bibitem{2011ApJ...733...66I}
Y.~{Inoue}, {\it {Contribution of Gamma-Ray-loud Radio Galaxies' Core Emissions
  to the Cosmic MeV and GeV Gamma-Ray Background Radiation}},  {\em \apj} {\bf
  733} (May, 2011) 66, [\href{http://xxx.lanl.gov/abs/1103.3946}{{\tt
  arXiv:1103.3946}}].

\bibitem{cosmowimp}
{\bf Fermi-LAT} Collaboration, M.~Ackermann et~al., {\it {Limits on Dark Matter
  Annihilation Signals from the Fermi LAT 4-year Measurement of the Isotropic
  Gamma-Ray Background}},  {\em JCAP} {\bf 1509} (2015), no.~09 008,
  [\href{http://xxx.lanl.gov/abs/1501.05464}{{\tt arXiv:1501.05464}}].

\bibitem{2012MNRAS.421L..87S}
P.~D. {Serpico}, E.~{Sefusatti}, M.~{Gustafsson}, and G.~{Zaharijas}, {\it
  {Extragalactic gamma-ray signal from dark matter annihilation: a power
  spectrum based computation}},  {\em \mnras} {\bf 421} (Mar., 2012) L87--L91,
  [\href{http://xxx.lanl.gov/abs/1109.0095}{{\tt arXiv:1109.0095}}].

\bibitem{2002PhR...372....1C}
A.~{Cooray} and R.~{Sheth}, {\it {Halo models of large scale structure}},  {\em
  \physrep} {\bf 372} (Dec., 2002) 1--129,
  [\href{http://xxx.lanl.gov/abs/astro-ph/0206508}{{\tt astro-ph/0206508}}].

\bibitem{Sefusatti:2014vha}
E.~Sefusatti, G.~Zaharijas, P.~D. Serpico, D.~Theurel, and M.~Gustafsson, {\it
  {Extragalactic gamma-ray signal from dark matter annihilation: an
  appraisal}},  {\em Mon.Not.Roy.Astron.Soc.} {\bf 441} (2014) 1861--1878,
  [\href{http://xxx.lanl.gov/abs/1401.2117}{{\tt arXiv:1401.2117}}].

\bibitem{2012MNRAS.422.3189G}
R.~C. {Gilmore}, R.~S. {Somerville}, J.~R. {Primack}, and
  A.~{Dom{\'{\i}}nguez}, {\it {Semi-analytic modelling of the extragalactic
  background light and consequences for extragalactic gamma-ray spectra}},
  {\em \mnras} {\bf 422} (June, 2012) 3189--3207,
  [\href{http://xxx.lanl.gov/abs/1104.0671}{{\tt arXiv:1104.0671}}].

\end{thebibliography}\endgroup


\providecommand{\href}[2]{#2}\begingroup\raggedright\endgroup

\end{document}